\PassOptionsToPackage{hypertexnames=false}{hyperref}

\documentclass[pdflatex,sn-mathphys-num]{sn-jnl}%

\usepackage{graphicx}%
\usepackage{multirow}%
\usepackage{amsmath,amssymb,amsfonts}%
\usepackage{amsthm}%
\usepackage{mathrsfs}%
\usepackage[title]{appendix}%
\usepackage{xcolor}%
\usepackage{textcomp}%
\usepackage{manyfoot}%
\usepackage{booktabs}%
\usepackage{algorithm}%
\usepackage{algorithmicx}%
\usepackage{algpseudocode}%
\usepackage{listings}%
\usepackage{multibib}
\newcites{si}{Supplementary References}
\usepackage{titletoc}
\usepackage{verbatim}

\usepackage{etoolbox}
\robustify{\cite}
\robustify{\citesi}

\usepackage{bm}

\newcommand{\bq}{\mathbf{q}}
\newcommand{\bS}{\mathbf{S}}

\newcommand{\Bsat}{B_{\mathrm{sat}}}    %
\newcommand{\Bc}{B_{\mathrm{flip}}}        %

\newcommand{\kb}{k_{\mathrm B}}
\newcommand{\TN}{T_{\mathrm N}}
\newcommand{\TC}{T_{\mathrm C}}
\newcommand{\q}{\mathbf q}

\DeclareMathOperator{\re}{Re}
\DeclareMathOperator{\im}{Im}

\theoremstyle{thmstyleone}%

\theoremstyle{thmstyletwo}%

\theoremstyle{thmstylethree}%

\begin{document}

\title[Optical observation of interlayer spin correlation]{
Optical observation of interlayer spin correlation
}

\author*[1]{\fnm{Akiyoshi}\sur{Park}}\email{akipark@umd.edu}
\author[1]{\fnm{Pranshoo} \sur{Upadhyay}}
\author[1]{\fnm{Andrey} \sur{Grankin}}
\author[2]{\fnm{Emil Vi\~{n}as} \sur{ Bostr\"{o}m}}
\author[1]{\fnm{Mahdi} \sur{Ghafariasl}}
\author[1]{\fnm{Hassan} \sur{Alnatah}}
\author[1]{\fnm{Masoud} \sur{Mohammadi-Arzanagh}}
\author[1]{\fnm{Gautam} \sur{Nambiar}}
\author[1]{\fnm{Beini} \sur{Gao}}
\author[1]{\fnm{Alireza} \sur{Alvandi}}
\author[1]{\fnm{Sakthi} \sur{Rajmano Madhan Kumar}}
\author[1]{\fnm{Ghadah} \sur{Alshalan}}
\author[1]{\fnm{Isaac} \sur{Sherwood}}
\author[1]{\fnm{Mahmoud} \sur{Jalali Mehrabad}}
\author[3]{\fnm{You} \sur{Zhou}}
\author[4]{\fnm{Arun} \sur{Ramanathan}}
\author[4]{\fnm{Xavier} \sur{Roy}}
\author[2,5]{\fnm{Angel} \sur{Rubio}}
\author*[1]{\fnm{Mohammad} \sur{Hafezi}}\email{hafezi@umd.edu}

\affil[1]{\orgdiv{Joint Quantum Institute}, \orgname{University of Maryland, College Park, Maryland 20742, USA}}

\affil[2]{\orgname{Max Planck Institute for the Structure and Dynamics of Matter, Luruper Chaussee 149, 22761 Hamburg, Germany}}

\affil[3]{\orgname{Department of Materials Science and Engineering}, \orgname{University of Maryland, College Park, Maryland 20742, USA}}

\affil[4]{\orgdiv{Department of Chemistry}, \orgname{Columbia University, New York, New York 10027, USA}}

\affil[5]{\orgdiv{Initiative for Computational Catalysis (ICC) and Center for Computational Quantum Physics (CCQ)}, \orgname{Flatiron Institute, New York, New York 10010, USA}}

\abstract{
Spin correlations govern numerous collective behavior of quantum materials, underpinning exotic phenomena such as unconventional superconductivity and topological magnetism. In layered materials, the interlayer spin correlation is important because it characterizes the magnetic ground state and determines spin transport across the interface. Yet interlayer spin correlations have remained hard to measure directly, leaving one of the most basic quantities of two-dimensional magnetism out of experimental reach. Here we provide the first direct optical probe of interlayer spin correlations, using two-magnon Raman scattering in the van der Waals antiferromagnet (AFM) CrSBr, in a remarkable agreement with a microscopic spin-wave model without any fitting parameter. Moreover, the two-magnon channel switches on only in the AFM state and vanishes when a magnetic field takes the crystal to a ferromagnetic state. We furthermore, establish an exciton-mediated variant of the technique, where tuning the laser near the exciton resonance enhances the signal roughly tenfold due to the exciton's large oscillator strength. Magnon-pair Raman spectroscopy thus opens a direct optical window into interlayer spin correlations in van der Waals magnets, extendable to twisted bilayers and proximity-coupled heterostructures.
}


\maketitle

\section*{Introduction}

Spin--spin correlations, $\langle\hat{\mathbf S}_{i}\cdot\hat{\mathbf S}_{j}\rangle$, are fundamental to the understanding of magnetism. They characterize the ground state and the emergent phenomena that arise from it, from unconventional superconductivity~\cite{RevModPhys.84.1383, doi:10.1126/science.235.4793.1196, RevModPhys.78.17, Keimer2015, Monthoux2007, RevModPhys.83.1589, RevModPhys.87.855} and quantum spin liquids to multiferroicity \cite{Kimura2003, Cheong2007, Song2022} and topological magnetism~\cite{Balents2010, Nagaosa2013, doi:10.1126/science.1166767, doi:10.1126/science.aay0668}. Measuring these correlations directly has thus been actively investigated \cite{DallaPiazza2015, RevModPhys.83.705, LeTacon2011, Kang2020}. A growing body of work has established optical access to magnetism in quantum materials, such as magneto-optical and second-harmonic probes of layer-dependent order in 2D magnetic materials ~\cite{Yang2020, Sun2019}, as well as magnetic circular dichroism spectroscopy in gate-controlled magnetic states in Moir\'e transition metal dichalcogenides (TMDs)~\cite{Tang2020, doi:10.1126/science.adg4268}. These probes, however, report only the ordered configuration or the single-spin order parameter $\langle\hat{\mathbf S}_{i}\rangle$.

Among the probes able to access them, Raman scattering stands out as a direct optical window onto spin--spin correlations~\cite{RevModPhys.79.175}. Yet from the classic antiferromagnetic  fluorides~\cite{PhysRevLett.18.658, PhysRevB.3.1709} to the correlated cuprates~\cite{PhysRevB.42.10220, PhysRevB.53.R11930}, magnetic Raman scattering has probed only the correlations {within} a single layer, governed by the dominant intralayer exchange. A far richer question lies {across} the layers.

In layered quantum materials, where two-dimensional confinement magnifies quantum fluctuations and the interlayer stacking sets the effective dimensionality, the interlayer correlator $\langle\hat{\mathbf S}_{A}\cdot\hat{\mathbf S}_{B}\rangle$ between adjacent layers $A$ and $B$ becomes the quantity of central interest. It distinguishes two- from three-dimensional order~\cite{Gibertini2019}, dictates whether spin currents traverse van der Waals (vdW) interfaces~\cite{doi:10.1126/science.aav4450, MacNeill2017}, and sets the proximity coupling that underpins magnetic heterostructures~\cite{Wei2016, Zhao2017}. Despite a growing body of theoretical work~\cite{PhysRevB.54.3468}, this correlator has remained experimentally elusive, with no direct optical measurement reported to date. Developing a spectroscopic channel that directly encodes $\langle\hat{\mathbf S}_{A}\cdot\hat{\mathbf S}_{B}\rangle$ is therefore a prerequisite for opening a path to coherent control of their collective spin degrees of freedom~\cite{67zs-hqf3}.

\begin{figure}[h!]
    \centering
    \includegraphics[width=\linewidth]{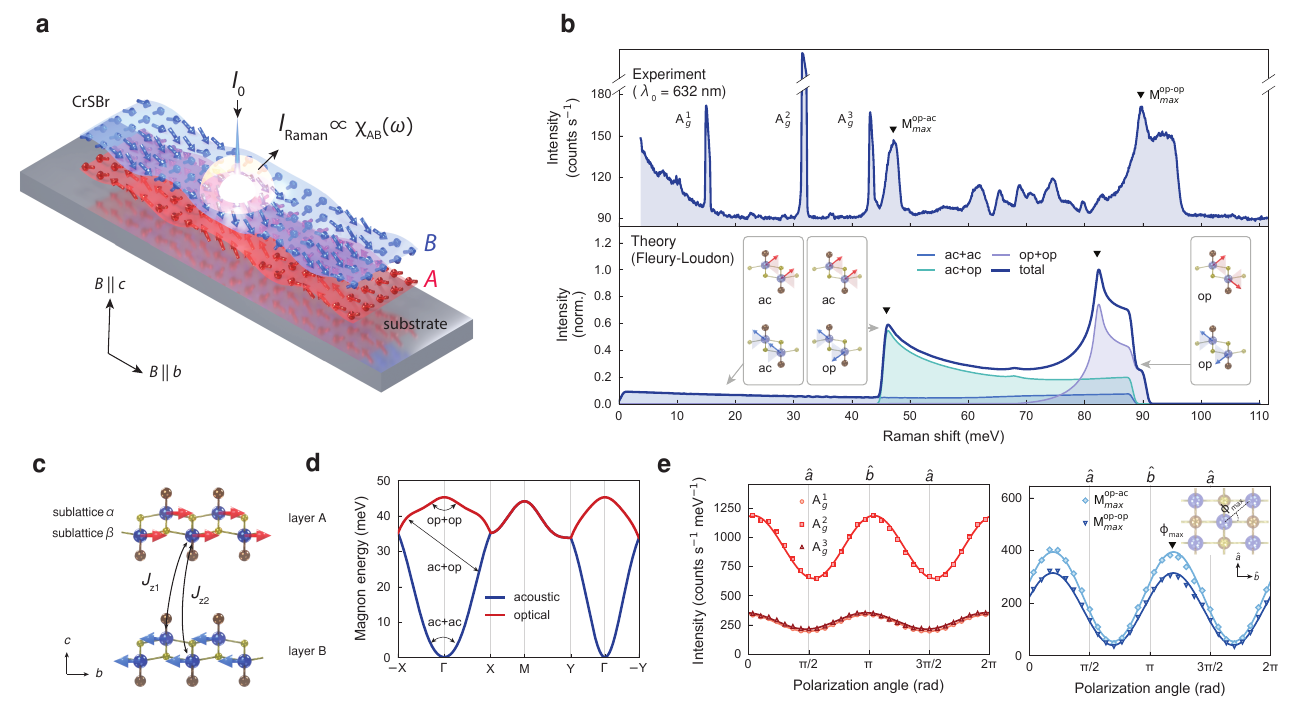}
    \caption{\textbf{Two-magnon Raman scattering in CrSBr.} \textbf{a}, Schematic of the Raman scattering setup used to measure the dynamical correlation function ($\chi_{AB}$) of the interlayer spin--spin operator $\hat{\mathbf S}_{A}\cdot\hat{\mathbf S}_{B}$ via two-magnon emission. \textbf{b}, Top, Experimental Raman spectrum of bulk CrSBr at $T = 3.5$~K, $\lambda_0 = 633$~nm, $P = 800~\mu$W. Bottom, Theoretical Fleury--Loudon two-magnon spectrum decomposed by parity channel, with $M_{\rm max}^{\rm op-ac}$ and $M_{\rm max}^{\rm op-op}$ labelled. \textbf{c}, Crystal structure along the $bc$-plane, with arrows marking the interlayer couplings $J_{z1}$ (diagonal) and $J_{z2}$ (axial). \textbf{d}, Magnon dispersion from intralayer exchanges obtained by inelastic neutron scattering (Table~\ref{tab:Jvalues})~\cite{https://doi.org/10.1002/advs.202202467}. \textbf{e}, Integrated Raman intensity versus in-plane polarization angle $\phi$ for the three $A_g$ phonons and the two magnon-derived bands ($M_{\rm max}^{\rm op-ac}$, $M_{\rm max}^{\rm op-op}$).}
    \label{fig:new_fig_1}
\end{figure}

We hereby present magnon-pair Raman scattering as an effective method of directly diagnosing the interlayer spin correlator in 2D vdW antiferromagnets (Fig.~\ref{fig:new_fig_1}(a)). Among two-dimensional magnets, the layered van der Waals semiconductor CrSBr stands out as an ideal testbed, as CrSBr is an optically active semiconductor that hosts excitons, and has shown strong exciton-magnon coupling \cite{Shao2025, Sun2024, Bae2022}. Moreover, it combines strong easy-axis anisotropy with ferromagnetic intralayer exchange, as well as weaker antiferromagnetic interlayer coupling (Fig.~\ref{fig:new_fig_1}(a)), producing a N\'{e}el order between adjacent layers \cite{https://doi.org/10.1002/adma.202003240, Lee2021, https://doi.org/10.1002/advs.202202467, Bo_2023, Esteras2022}. An applied magnetic field cants the localized spins of the Cr atoms away from the magnetic easy axis, overcoming the interlayer magnetic exchange \cite{Wilson2021}. Hence, the presence of interlayer exchange, together with the capability of actively controlling $\langle\hat{\mathbf S}_{A}\cdot\hat{\mathbf S}_{B}\rangle$ by applying a magnetic field makes CrSBr an excellent platform on which to search for an optical fingerprint of the interlayer spin correlator.

\section*{Main}

\subsection*{Two-magnon emission and the Fleury--Loudon continuum}\label{sec:FL}

We drive a variety of spontaneous Raman processes by exciting bulk CrSBr above the optical band gap, producing the spectrum in Fig.~\ref{fig:new_fig_1}(b). The sharp peaks at low Raman shift correspond to the three previously reported phonon modes~\cite{PhysRevB.107.075421, PhysRevB.111.174434, https://doi.org/10.1002/adfm.202211366, Mondal2025, Sahu2025}, as well as a continuum of features  between $\sim 45$ and $\sim 100$~meV, which has not been investigated before, deviating from the Lorentzian line shape.
Their strikingly broad lineshape is an indication of a two-particle emission process \cite{PhysRevB.49.8764, PhysRevB.59.7282}. Unlike single-particle scattering, which is pinned near zero momentum, a two-particle process emits its two quanta with equal and opposite momenta drawn from across the entire Brillouin zone. The spectrum therefore abandons the discrete Lorentzians of one-particle Raman for a broad continuum weighted by the joint density of states (JDOS). Decisively, these broad features vanish above the Curie temperature $T_\text{C} \sim 150$~K~\cite{https://doi.org/10.1002/adma.202204940}, pinning their origin as magnetic (Supplementary Information~\ref{SI_Tdependence}).

To account for the observed continuum, we model the magnetism of CrSBr with the spin Hamiltonian $\sum_{\langle ij\rangle} J_{ij}\,\hat{\mathbf S}_{i}\cdot\hat{\mathbf S}_{j}$, which sets the magnetic ground state, with $J_{ij}$ the Heisenberg exchange constant on bond $\langle ij\rangle$. The Raman scattering is driven by the Fleury--Loudon (FL) light--matter interaction $\sum_{\langle ij\rangle} \Lambda_{ij}\,J_{ij}\,\hat{\mathbf S}_{i}\cdot\hat{\mathbf S}_{j}$~\cite{PhysRev.166.514, PhysRevLett.24.1346}, whose polarization prefactor $\Lambda_{ij} = \bigl(\hat{\mathbf e}_{\rm in}\cdot\hat{\mathbf d}_{ij}\bigr)\bigl(\hat{\mathbf e}_{\rm out}^{*}\cdot\hat{\mathbf d}_{ij}\bigr)$ projects the incoming and outgoing photon polarizations onto the bond direction $\hat{\mathbf d}_{ij}$. In our measurements the light is polarized in the crystalline $a$--$b$ plane, so this factor keeps only the in-plane projection of each bond. This has a decisive consequence for what we detect. The two-magnon signal is carried entirely by the interlayer antiferromagnetic bonds, creating one magnon on each layer, and of the two interlayer couplings, $J_{z1}$ (diagonal) and $J_{z2}$ (axial) (Fig.~\ref{fig:new_fig_1}(c))~\cite{Bo_2023}, the axial $J_{z2}$ lies along $\hat c$ with no in-plane projection, so it is optically dark and the diagonal $J_{z1}$ alone supplies the Raman amplitude (Supplementary Information~\ref{SI_magnon} and~\ref{SI_FL}).
A Fermi-golden-rule treatment of $\hat{H}_\text{FL}$ relates the measured intensity to the dynamical correlation function ($\chi_{AB} (\omega)$ ) of the interlayer bond operator $\hat{\mathbf S}_{A}\cdot\hat{\mathbf S}_{B}$ \cite{RevModPhys.79.175},
\begin{align}
I_\text{Raman}(\omega) &\;\propto\;  \int dt\, e^{i\omega t}\,\big\langle (\hat{\mathbf S}_{A}\cdot\hat{\mathbf S}_{B})(t)\,(\hat{\mathbf S}_{A}\cdot\hat{\mathbf S}_{B})(0)\big\rangle \equiv {\chi}_{AB} (\omega) \notag \\
&\;\propto\; \sum_{\bq}\,|F(\bq)|^{2}\,\delta\bigl(\omega-\omega^{(\nu_{1})}_{\bq}-\omega^{(\nu_{2})}_{-\bq}\bigr).
\label{eq:I2m}
\end{align}
Because the scattering operator is the interlayer bond itself, $I_\text{Raman}$ is the dynamical autocorrelation of $\hat{\mathbf S}_{A}\cdot\hat{\mathbf S}_{B}$, a four-spin fluctuation spectrum whose weight is set by the interlayer exchange and by how strongly adjacent layers remain anti-aligned. It is this integrated weight, rather than the static expectation value $\langle\hat{\mathbf S}_{A}\cdot\hat{\mathbf S}_{B}\rangle$ itself, that the two-magnon intensity reports, so the signal collapses as the interlayer order is lost. Here $\omega^{(\nu)}_{\bq}$ is the branch-$\nu$ magnon energy and $F(\bq)=\sum_{\langle ij\rangle\,\in\,J_{z1}}\Lambda_{ij}\,J_{z1}\,\cos(\bq\cdot\hat{\mathbf d}^{\,\parallel}_{ij})$ is the FL form factor over the diagonal $J_{z1}$ bonds, with $\hat{\mathbf d}^{\,\parallel}_{ij}$ the in-plane bond projection. The right-hand form is the two-magnon JDOS weighted by $|F(\bq)|^{2}$. Momentum and spin conservation emit the two magnons at opposite momenta $\pm\bq$ with opposite $\Delta S^{z}=\pm1$ along $\hat b$ (Fig.~\ref{fig:new_fig_1}(c)), so van Hove singularities of the JDOS set the Raman lineshape.

As shown in Fig.~\ref{fig:new_fig_1}(b), the measured two-magnon spectrum is dominated by two pronounced features, a band edge near $46$~meV and a stronger peak near $82$~meV, both of which emerge naturally from the FL two-magnon continuum. Because CrSBr has two Cr sublattices per layer, the magnon dispersion carries two branches (Fig.~\ref{fig:new_fig_1}(d)), a lower-energy acoustic mode (${\rm ac}$) in which the sublattice spins precess in phase and a higher-energy optical mode (${\rm op}$) in which they precess out of phase. The two-magnon continuum therefore splits into three channels (${\rm ac}+{\rm ac}$, ${\rm ac}+{\rm op}$, ${\rm op}+{\rm op}$) (Fig.~\ref{fig:new_fig_1}(b), bottom and Fig. \ref{fig:SI_twomagnon}), making it far richer than a simple convolution of the single-magnon density of states. The model assigns the two observed features to van Hove singularities of these channels. The $46$~meV band edge $M^{\rm op-ac}_{\rm max}$ arises from the optical magnon gap at $\Gamma$ in the ${\rm ac}+{\rm op}$ channel, and the dominant $82$~meV peak $M^{\rm op-op}_{\rm max}$ from the $M$-corner saddle of $2\omega_{\rm op}(\bq)$ in the ${\rm op}+{\rm op}$ channel (Supplementary Information~\ref{SI_twomagnon}). The computed lineshape captures the full measured spectrum (Fig.~\ref{fig:new_fig_1}(b)), with both predicted peaks clearly resolved in the data. Aside from a $\sim 10$~meV shift of the $M^{\rm op-op}_{\rm max}$ peak, overall, the FL framework quantitatively accounts for the measured two-magnon peaks. We attribute this residual blueshift to magnon-magnon interactions, which a ladder/RPA resummation of the quartic spin-wave terms recasts as a repulsive interaction that pushes the optical-optical pair into an antibound state above the bare continuum, shifting the peak upward by a projected interaction $U_R \sim 10$~meV (See Supplementary Information \ref{SI_twomagnon_interaction}) \cite{andrei2026universalmagneticenergyscale}. The effect is far stronger for $M^{\rm op-op}_{\rm max}$ than for the acoustic-optical channel because the nearly flat optical band yields a large two-magnon density of states that lets repeated scattering build a well-defined antibound resonance, whereas the dispersive acoustic-optical continuum washes it out (Fig. \ref{fig:SI_twomagnon}).

Furthermore, the polarization dependence provides a decisive fingerprint of the magnetic origin. We mapped the emission of both the phonons and the two-magnon continuum  by varying the light polarization, where we observe that the $A_g$ phonons peak along the $\hat b$ axis, as dictated by the lattice. On the other hand, remarkably the two-magnon continuum behaves entirely differently, peaking at $\phi_{\max}\approx\arctan(b/a)$ from $\hat a$ (Fig.~\ref{fig:new_fig_1}(e), Fig.~\ref{fig:SI_pol}; Supplementary Information~\ref{SI_pol}), aligned with the in-plane projection of the interlayer $J_{z1}$ bonds, which runs along the same diagonal as the dominant intralayer $J_2$ exchange bond (Table~\ref{tab:Jvalues}). This is exactly the bond-projection response of the FL vertex, so the emission follows the interlayer magnetic bond geometry rather than the crystal lattice, confirming that it originates from magnon pairs rather than from phonons. The same signature places the active spin projection along the dominant $J_2$ direction, $\pm\phi_{\max}$. These two orientations are mirror-equivalent and degenerate, so equal populations of $+\phi_{\max}$ and $-\phi_{\max}$ domains would average to a net moment along $\hat b$, recovering the conventional macroscopic picture. Instead, we observe only $+\phi_{\max}$, consistent with our measurement probing a single magnetic domain.

\subsection*{Closing the interlayer AFM channel}

\begin{figure}[h!]
    \centering
    \includegraphics[width=\linewidth]{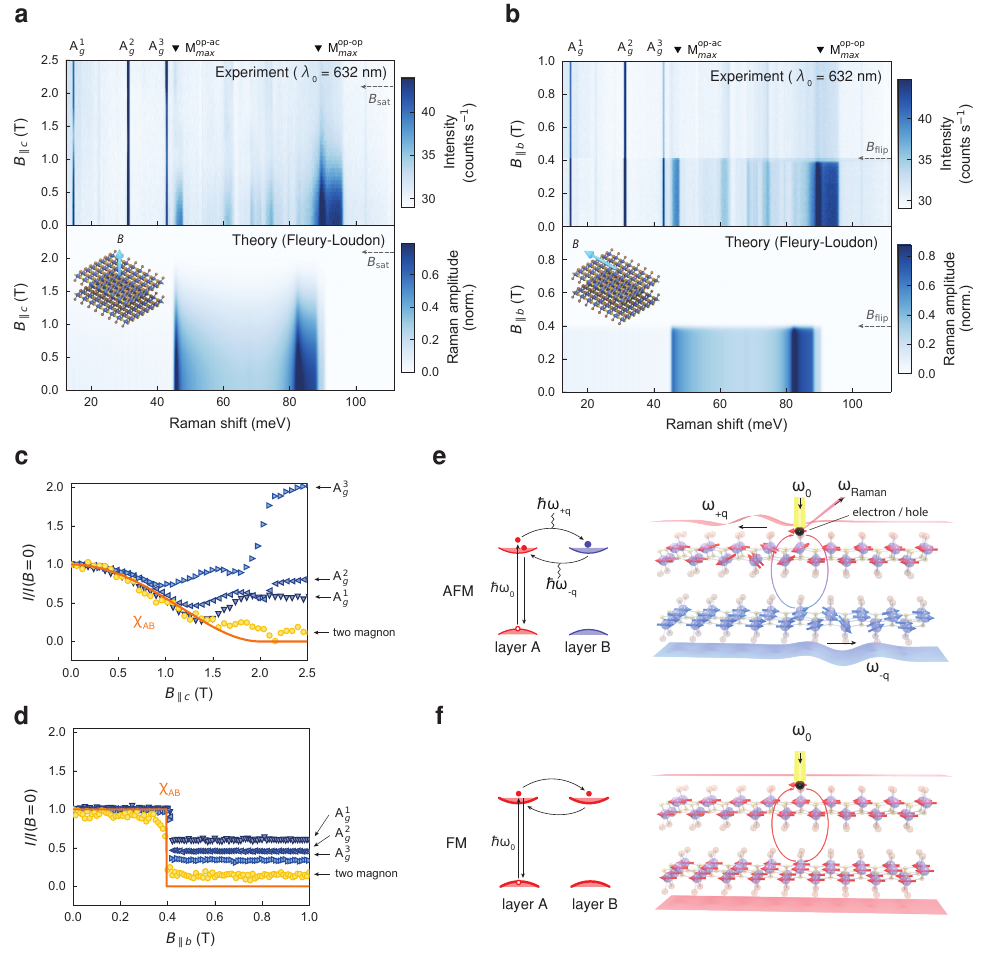}
    \caption{\textbf{Magnetic-field control of the CrSBr two-magnon.} \textbf{a}, Hard-axis sweep ($B \parallel c$). Top, measured Raman intensity ($P = 50~\mu$W, $T = 3.5$~K, $\lambda_0 = 633$~nm). The two-magnon collapses smoothly up to $B_{\rm sat}$. Bottom, Fleury--Loudon spectrum weighted by $\chi_{AB}$. \textbf{b}, Easy-axis sweep ($B \parallel b$). Top, the two-magnon is abruptly extinguished at the spin-flip field $B_{\rm flip} \approx 0.40$~T. Bottom, same vertex with the amplitude modulated by a tanh step at $\Bc$. \textbf{c},\textbf{d}, Integrated intensity versus $B \parallel c$ (\textbf{c}) and $B \parallel b$ (\textbf{d}) for three phonon bands (blue) and the two two-magnon peaks (gold), normalized to $B=0$. Solid orange lines, $(1-{B^{2}}/{(B_\text{sat})^{2}})^{2}$ fit in panel (\textbf{c}) and $\Theta(B_\text{flip}-B)$ fit for panel (\textbf{d}). \textbf{e}, AFM bilayer. The photo-excited electron tunnels between layers and emits a magnon at each hop. \textbf{f}, Field-polarized bilayer above $\Bsat$ or $\Bc$. The electron/holes delocalizes across both layers, but its projection onto the magnon-emitting anti-aligned sector collapses to zero, so the two-magnon channel is closed.}
    \label{fig:fig_2}
\end{figure}

The magnetic field gives us a knob to tune the interlayer correlation $\langle\hat{\mathbf S}_{A}\cdot\hat{\mathbf S}_{B}\rangle$ directly. Photoluminescence tracks how the interlayer order evolves with field (Fig.~\ref{fig:SI_PL}(a,b); Supplementary Information~\ref{SI_PL}). For $\mathbf B \parallel \hat c$ (hard axis), the order evolves continuously and reaches a forced ferromagnetic (FM) state at the saturation field $\Bsat = 2.1$~T. For $\mathbf B \parallel \hat b$ (easy axis), it switches abruptly to the FM state through a first-order spin-flip transition at $B_{\rm flip} \approx 0.4$~T. These behaviors correspond, respectively, to a smooth canting of the sublattice moments toward the field and to a discontinuous flip of one sublattice.

Under an out-of-plane field $\mathbf B \parallel \hat c$, the two-magnon peaks collapse dramatically, vanishing by $B \sim 2$~T, being consistent with the saturation field $\Bsat = 2.1$~T at which the interlayer spins fully align. The phonon lines, by contrast, barely move, changing in intensity by only $\sim 20$--$50\%$ and never vanishing (Fig.~\ref{fig:fig_2}(a, top; c)). This striking selectivity is a direct consequence of the FL mechanism. The two-magnon amplitude is controlled at the matrix-element level by the interlayer spin correlator, so its intensity is forced to zero as the layers cant into a common spin configuration at $\Bsat$. The phonons carry no spin-dependent matrix element and respond to the magnetic state only weakly and indirectly, through the resonance condition and possible magneto-elastic coupling.

The collapse follows a well-defined envelope. Canting the moments by $\theta(B)$ away from the $\hat b$ easy axis rescales the interlayer magnon-pair-creation amplitude by $\cos^{2}\theta$, which tracks how anti-aligned the two layers remain (Supplementary Information~\ref{SI_Bfield}). With $\sin\theta = B/\Bsat$, so $\cos^{2}\theta = 1-(B/\Bsat)^{2}$, the two-magnon response $\chi_{AB}(\omega)$ is scaled by an overall $\cos^{4}\theta = [1-(B/\Bsat)^{2}]^{2}$ envelope that vanishes as the moments fully polarize at $\Bsat$. Remarkably, the FL spectrum computed with this envelope reproduces the measured collapse (Fig.~\ref{fig:fig_2}(a), bottom panel). The small residual intensity that survives above the transition, which the two-magnon mechanism alone cannot account for, is a weak, field-independent channel mediated by spin-orbit coupling, which mixes the up- and down-spin states and relaxes the strict interlayer selection rule.
Intriguingly, the two-magnon peak energies barely shift with field (Supplementary Information~\ref{SI_magnonBfield}), a direct signature that the magnon dispersion is essentially unchanged. Only the anisotropy gap and Zeeman terms move the energies, by ${\sim}\,10^{-2}$~meV, far below our experimental resolution.

A sharp quench appears for an in-plane field $\mathbf B \parallel \hat b$. The two-magnon band is nearly unchanged below $B_\text{flip}$ but drops abruptly by ${\sim}\,90\%$ across the spin-flip, while the phonons are unaffected (Fig.~\ref{fig:fig_2}(b)). This step-like collapse is exactly the first-order response predicted by the FL framework, in contrast to the smooth $\cos^{4}\theta$ decay along $\hat c$. That the two-magnon emission dies in the spin-polarized state under both field geometries, precisely when the interlayer antiferromagnetic order is destroyed, establishes it as a faithful all-optical probe of the interlayer spin correlations. The signal nonetheless persists in temperature up to $T_\text{C}$ rather than $T_\text{N}$, because it requires intralayer magnons and reflects the dynamical interlayer correlator, which retains short-range weight above $T_\text{N}$ even after long-range interlayer order is lost.

To pin down the microscopic origin, we trace the interlayer FL term $J_{z1}\hat{\mathbf S}_A\cdot\hat{\mathbf S}_B$ to a virtual charge-transfer loop between the layers (Fig.~\ref{fig:fig_2}(e,f)). In the AFM ground state (Fig.~\ref{fig:fig_2}(e)), an incoming photon $\hbar\omega_0$ virtually promotes an electron on layer $A$, which hops to layer $B$ across the antiparallel spin background and emits one magnon on each layer, at $\hbar\omega_{+\mathbf q}$ and $\hbar\omega_{-\mathbf q}$, before radiative recombination. This loop generates the magnon pair that carries the two-magnon Raman signal. In the field-polarized FM state (Fig.~\ref{fig:fig_2}(f)) the two layers share the same spin polarization, so the interlayer hop is Pauli-blocked and the two-magnon channel closes, exactly the quench observed at high field

\subsection*{Exciton-mediated two-magnon emission}\label{sec:KH_main}

\begin{figure}[h!]
    \centering
    \includegraphics[width=\linewidth]{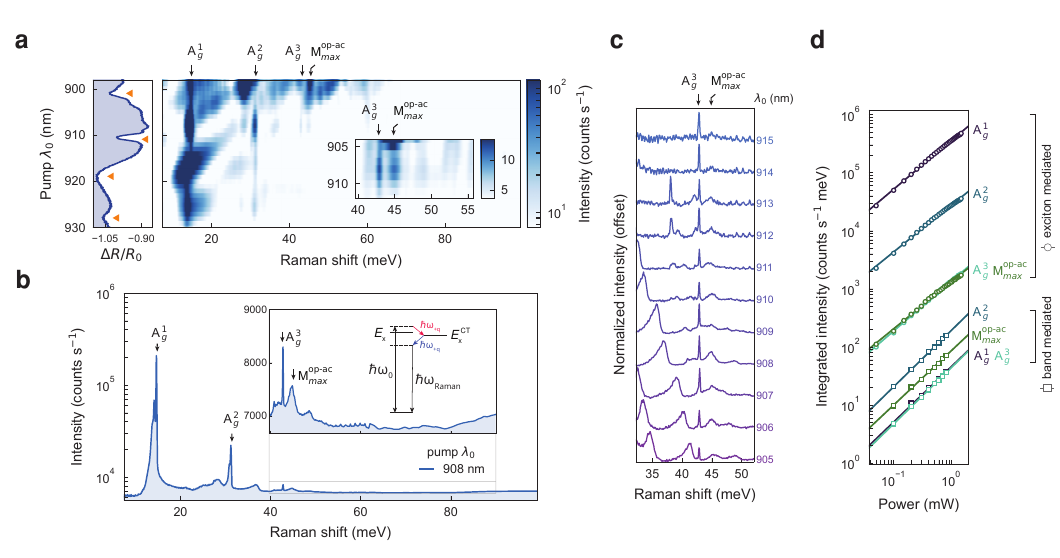}
    \caption{\textbf{Raman scattering mediated by excitonic states in CrSBr.} \textbf{a}, Excitation-wavelength dependence of the differential reflectivity (left) and of the Raman spectrum (right) over 900 to 930~nm. Inset, zoom on the 908~nm exciton resonance where the two-magnon features are visible ($T = 3.5$~K, $P = 500~\mu$W). \textbf{b}, Raman spectra at $\lambda_0 = 908$~nm, inset; An energy diagram depicting the exciton (with energy $E_\text{X}$) mediated process to emit two-magnons, going through a dipole forbidden charge-transfer (CT) exciton intermediate with energy, $E_\text{X}^\text{CT}$. \textbf{c}, Waterfall plot of the excitation-wavelength dependence of the Raman emission in the two-magnon window. \textbf{d}, Power dependence of the integrated Raman intensity for the principal phonon and magnon features at near-resonant (908~nm) and off-resonant (633~nm) excitation.}
    \label{fig:fig_3}
\end{figure}

The interlayer two-magnon vertex is intrinsically weak, so to bring it out we drive the Raman process on resonance with CrSBr's excitons, which concentrate oscillator strength and couple strongly to the spins~\cite{Dirnberger2026, Diederich2023, Datta2025}. Whereas our earlier spectra were excited across interband electronic transitions, near-resonant excitation at the exciton sharply enhances the two-magnon signal (Fig.~\ref{fig:fig_3}(a), right panel). Using differential reflectivity, we first identify the exciton peaks and their satellites, which reflect exciton-phonon~\cite{Lin2024, Śmiertka2026} and exciton-polariton dressing~\cite{Dirnberger2023, Wang2023} (Fig.~\ref{fig:fig_3}(a)). As we scan the excitation wavelength across these resonances, the $A_g$ phonons sharpen near the absorption peaks and the two-magnon feature stands clear of the PL background. The Raman peaks stay fixed in Raman shift as $\lambda_0$ varies while the PL features move with $\lambda_0$ (Fig.~\ref{fig:fig_3}(a,c)), cleanly separating the two processes.

Strikingly, the exciton-mediated and band-mediated spectra differ even though they probe the same magnetic excitations, with the exciton-mediated spectra suppressing the higher-energy Raman features. This decay with increasing Raman shift reflects the stringent resonance condition set by the narrow exciton line, in contrast to the broad electronic continuum of band-mediated Raman. As the Raman shift $\omega$ grows, the emitted photon drifts off the exciton line and the dipole matrix element falls off.

The laser-power dependence of the Raman scattering quantifies this intensity  enhancement by mediating excitons. The phonon and magnon peaks are an order of magnitude stronger under exciton-mediated excitation than under band-mediated excitation (Fig.~\ref{fig:fig_3}(d), Figs.~\ref{fig:SI_power_1} and \ref{fig:SI_power_2}; Supplementary Information~\ref{SI_power}). This amplification arises because the exciton concentrates oscillator strength into a single bound state, whereas the off-resonant response samples a continuum of band states with no single resonant pole to lock onto. Harnessing the exciton as a resonant amplifier turns an otherwise faint magnon-pair signal into a robust probe, a strategy that should extend across the broader family of optically active 2D magnets.

\begin{figure}[h!]
    \centering
    \includegraphics[width=0.7\linewidth]{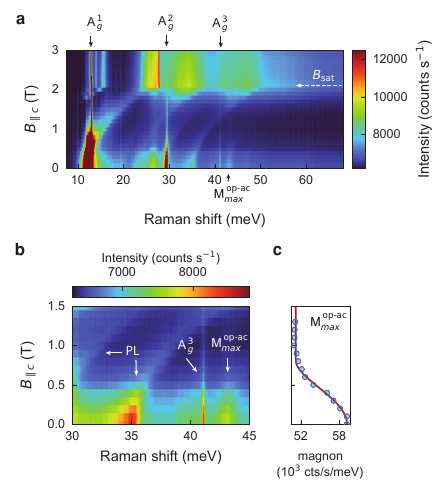}
    \caption{\textbf{Magnetic-field control of the exciton-mediated two-magnon emission in CrSBr.} \textbf{a}, Raman spectra as a function of magnetic field $\mathbf{B} \parallel \hat c$ at $T = 3.5$~K, $P = 500\,\mu$W, and $\lambda_0 = 908$~nm. \textbf{b}, zoom on the two-magnon spectral window of (a), highlighting the optical-magnon band edge $M^{\rm op-ac}_{\rm max}$. \textbf{c}, Integrated intensity of $M^{\rm op-ac}_{\rm max}$ versus $B$, with the quartic canting envelope (solid line).}
    \label{fig:fig_4}
\end{figure}

Under an out-of-plane field $\mathbf B \parallel \hat c$, the Raman peaks stay fixed in energy while the PL features redshift continuously (Fig.~\ref{fig:fig_4}(b)), as spin-conserving interlayer tunnelling turns on and the exciton delocalizes between layers~\cite{Wilson2021, Shao2025}. The two-magnon emission $M^{\rm op-ac}_{\rm max}$ weakens with field and vanishes while the phonon lines remain intact, mirroring the electronic-band-mediated case (Fig.~\ref{fig:fig_2}). The magnetic response is thus identical whichever channel we excite through, confirming the magnon origin of the signal. The peak diminishes before $\Bsat$ because the field shifts the exciton energy and increases the detuning, further suppressing the resonant Raman intensity.

Carrying the two-magnon amplitude between the layers requires a virtual intermediate that spans both layers, and in the exciton-mediated process this is an interlayer charge-transfer (CT) exciton, with an electron on one layer bound to a hole on an adjacent layer (Fig. \ref{fig:fig_4}(b) inset). Interlayer CT excitons in CrSBr have so far only been predicted theoretically~\cite{PhysRevB.111.075107, PhysRevB.111.205301}. Because such an exciton is the natural virtual intermediate for the interlayer process, the exciton-mediated two-magnon emission is consistent with, and most naturally explained by, the presence of an interlayer CT exciton.

\section*{Outlook}

Through this investigation, we use have used an optical technique to measure interlayer two-magnon emission through a Raman process. Beyond serving as a probe, a coherent Raman drive can be used to optically manipulate spin correlations, thereby engineering magnetic states inaccessible in equilibrium  ~\cite{Claassen2017, Vogl_2023, Fadler_2024, huang2025opticalengineeringdetectionmagnetism}. A further direction is to move from photon intensities to their quantum correlations \cite{Mentink2015, Bloch2022, 67zs-hqf3}. While the two-magnon intensity measured here is the dynamical four-spin correlator of the interlayer bond operator, by correlating the Stokes and anti-Stokes scattered photons, one can access the higher-order correlations between magnon pairs beyond linear response. Because the Raman scattering process imprints the interlayer bond operator onto the emitted photon, the scattered photons physically inherit the spin correlations of the material, so the underlying spin-spin correlations appear as genuine correlations between the scattered photons that can be read out with Hanbury Brown--Twiss and homodyne detection~\cite{67zs-hqf3}.

Moreover, this strategy extends well beyond CrSBr and applies to magnetic heterostructures. Since the interlayer signal scales with the number of interlayer bonds, reaching the atomically thin limit is demanding. For example, in twisted bilayers and proximity-induced heterostructures, the interlayer correlation is set by the stacking angle~\cite{Tong2018, Mak2019} and exchange induced across the interface~\cite{PhysRevLett.124.197401}, so the same measurement directly reads out how neighboring spins orient relative to one another. This further becomes a particularly powerful technique when an optically active magnet is interfaced with an exotic state such as a superconductor or a topological material, where spin correlations can be optically measured.

Finally, the remarkable parameter-free agreement between our measurements and the microscopic Fleury--Loudon spin-wave model pins down the light--magnon coupling that governs the two-magnon process, providing a foundation for optically probing, controlling, and engineering interlayer spin correlations in van der Waals magnets.

\section*{Methods}
\subsection*{Sample preparation}
Bulk CrSBr single crystals were grown by chemical vapor transport using stoichiometric quantities of the elemental precursors, following the procedure detailed in Ref.~\cite{https://doi.org/10.1002/adma.202003240}. Bulk-like flakes were mechanically exfoliated from the parent crystals using an adhesive tape and transferred onto pre-cleaned Si substrates capped with a 285~nm thermally grown SiO$_2$ layer. Flake thickness of 50 nm was characterized by atomic force microscopy. All samples were kept under an inert atmosphere between exfoliation and cryostat loading to minimize environmental degradation of the freshly cleaved surfaces.

\subsection*{Optical measurements}
All optical measurements were carried out in a closed-cycle He cryostat with an integrated superconducting vector magnet, with the sample held at $T = 3.5$~K. Photoluminescence spectra were excited by a $633$~nm He-Ne laser (Thorlabs) cleaned by a $1$~nm bandpass filter, with the collected emission dispersed by a spectrometer (Princeton Instruments) equipped with a $1200$~lines/mm grating. Differential reflectivity was measured with a supercontinuum white-light source (NKT Photonics), reflectivity being defined as the difference between the light reflected from the sample and from the bare substrate normalized to the substrate reflectance. Off-resonant Raman spectra were acquired with the same $633$~nm He-Ne laser, with the excitation polarization set by a linear polarizer followed by a rotatable half-wave plate, the incident power controlled by a variable neutral-density (ND) filter, the Rayleigh line rejected by a notch filter (OptiGrate), and the Stokes signal dispersed on the same spectrometer. Near-resonant Raman spectra were acquired with a continuous-wave tunable Ti:sapphire laser (M Squared Lasers) scanned across the intralayer exciton at $\sim$908~nm, with the Rayleigh line rejected by a tunable  long-pass filter (Semrock) matched to each excitation wavelength. Power drift during the wavelength sweep was eliminated by actively stabilizing the incident power through a closed feedback loop between a variable ND filter and an in-line power meter.

\subsection*{Numerical calculations}

First-principles calculations were performed with plane-wave DFT in Quantum ESPRESSO~\cite{Giannozzi_2017}, using the experimental bulk orthorhombic structure of CrSBr from Ref.~\cite{https://doi.org/10.1002/adma.202003240} (space group $Pmmn$, $a = 3.55335$~\AA, $b = 4.74493$~\AA, $c = 17.521$~\AA). The unit cell contains two CrSBr layers stacked along $\hat c$, each with two crystallographically distinct Cr sublattices. Exchange and correlation were treated with PBE plus on-site Hubbard corrections $U_{\rm eff} = 3$~eV on the Cr $3d$ manifold in the ortho-atomic scheme, with plane-wave cutoffs $E_{\rm cut}^{\rm wfc} = 100$~Ry and $E_{\rm cut}^{\rm rho} = 800$~Ry converged on a $16 \times 12 \times 1$ Monkhorst-Pack $k$-mesh to $10^{-7}$~Ry. Two magnetic ground states were computed with collinear spin polarization and no spin-orbit coupling. The AFM state with opposite magnetization on the two layers, reproducing the experimental A-type interlayer N\'{e}el order, and the FM state with parallel magnetization on all four Cr sites.

A non-self-consistent (NSCF) step on the same $k$-mesh with $148$ bands provided the Bloch states for the maximally localized Wannier construction in Wannier90~\cite{Pizzi_2020}. The Wannier basis comprised $44$ orbitals per spin channel. Five $3d$ orbitals ($d_{z^2}$, $d_{xz}$, $d_{yz}$, $d_{x^2-y^2}$, $d_{xy}$) on each Cr site, and three $p$ orbitals on each S and Br site. From the resulting Wannier Hamiltonians we computed the spin- and orbital-resolved band structure and projected the eigenstates onto the Cr-$d$ + S-$p$ subspace of each layer to produce the fat-band plots of Fig.~\ref{fig:new_fig_1}. The Br-$p$ orbitals are retained in the basis to preserve the DFT band structure but are excluded from the reported weights, as they contribute negligibly to the low-energy valence and conduction states. To compensate for the well-known PBE+U underestimation of the semiconducting gap, we applied a rigid scissor correction $\Delta$ to all conduction states ($E > E_F$) chosen to open the AFM gap to the experimental optical/QP value of $1.50$~eV. The same $\Delta$ was applied to the FM band structure for consistency.

\section*{Acknowledgements}
Y.Z.\ acknowledges support from the U.S.\ Department of Energy, Office of Science,
This work was supported by the Max Planck–New York City Center for Non-Equilibrium Quantum Phenomena and by the Cluster of Excellence Advanced Imaging of Matter (AIM). The Flatiron Institute is a division of the Simons Foundation.
Synthesis work at Columbia University was supported by the Department of Energy under award number DE-SC0023406 (X.R.).
Office of Basic Energy Sciences, Award No.\ DE-SC0022885.
Funding for the AFM shared facility used in this research was provided by NSF under
award number CHE-1626288.

\section*{Author Information}

\subsection*{Author contributions}
A.P., P.U., A.G.\ and M.H.\ conceived the project. A.P.\ and P.U.\ performed the optical
measurements with assistance from M.G., H.A.\ and B.G. A.Ram., G.A.\ and I.S.\ prepared and
characterized the samples. A.P., A.G., E.V.B., M.M.A., G.N., A.A.\ and S.R.M.K.\ developed
the theoretical model and performed the numerical calculations. A.P.\ analyzed the data
with input from all authors. A.P., P.U., A.G., E.V.B., M.M.A. \ and M.H.\ wrote the manuscript with
contributions from all authors. X.R.\ supervised the sample synthesis, A.Rub.\ supervised
the theoretical work, and M.H.\ supervised the project. All authors discussed the results
and commented on the manuscript.

\subsection*{Competing interests}
The authors declare no competing interests.

\subsection*{Data availability}
The data supporting the findings of this study are presented in the paper and its
Supplementary Information. The raw spectroscopy and magnetic-field datasets are available
from the corresponding author upon reasonable request.

\subsection*{Code availability}
The custom code used to compute the magnon dispersion, the Fleury--Loudon two-magnon
spectra, and the RPA magnon--magnon corrections is available from the corresponding author
upon reasonable request. The first-principles calculations were performed with the publicly
available Quantum ESPRESSO and Wannier90 packages.

\bibliography{sn-bibliography-main}

\newpage
\begin{center}
{\Large\textbf{Supplementary Information for}}\\[6pt]
{\Large\textbf{Optical observation of interlayer spin correlation}}\\[10pt]
{\normalsize
Akiyoshi Park, Pranshoo Upadhyay, Andrey Grankin, Emil Viñas Boström, Mahdi Ghafariasl, Hassan Alnatah, Masoud Mohammadi-Arzanagh, Gautam Nambiar, Beini Gao, Alireza Alvandi, Sakthi Rajmano Madhan Kumar, Ghadah Alshalan, Isaac Sherwood, Mahmoud Jalali Mehrabad, You Zhou, Arun Ramanathan, Xavier Roy, Angel Rubio, and Mohammad Hafezi.
}\\[6pt]
\end{center}
\vspace{6pt}
\setcounter{section}{0}
\setcounter{equation}{0}
\setcounter{figure}{0}
\setcounter{table}{0}
\renewcommand{\thesection}{S\arabic{section}}
\renewcommand{\theequation}{S\arabic{equation}}
\renewcommand{\thefigure}{S\arabic{figure}}
\renewcommand{\thetable}{S\arabic{table}}
\renewcommand{\theHsection}{SI.\arabic{section}}
\renewcommand{\theHequation}{SI.\arabic{equation}}
\renewcommand{\theHfigure}{SI.\arabic{figure}}
\renewcommand{\theHtable}{SI.\arabic{table}}

\startcontents
\section*{Contents}
\printcontents{}{1}{\setcounter{tocdepth}{2}}

\section{Temperature dependence of the Raman spectrum}\label{SI_Tdependence}
\begin{figure}[h!]
    \centering
    \includegraphics[width=\linewidth]{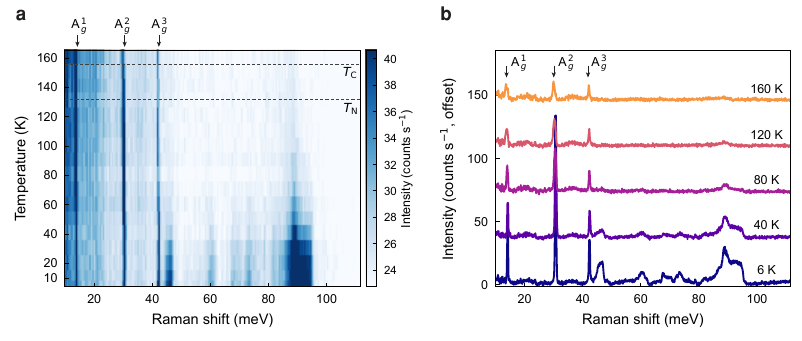}
    \caption{\textbf{Temperature dependence of Raman Spectrum.}
    \textbf{a}, A 2D map exhibiting the Raman spectra taken from 4 K to 160 K. \textbf{b}, A waterfall plot of selected temperatures of the Raman spectra. The temperature dependence was measured with $\lambda_0 = $ 633 nm and $P = 100$ $\mu$W.}
    \label{fig:SI_Tdependence}
\end{figure}

The temperature dependence of Raman spectroscopy taken from $T = 6$ K to 160 K is displayed in Fig.~\ref{fig:SI_Tdependence}. Below $T_\text{C}\!=\!150$~K, the intralayer Cr moments align ferromagnetically along the easy axis $\hat b$ and below $T_\text{N}\!=\!132$~K, the two layers couple antiferromagnetically, producing the interlayer N\'{e}el ground state ~\citesi{Lee2021-si,https://doi.org/10.1002/adma.202003240-si}.

\section{Magnon structure}\label{SI_magnon}
\subsection{Heisenberg spin Hamiltonian}

The magnon spectrum of bulk CrSBr is captured by a bilayer Heisenberg model with single-ion easy-axis anisotropy and a Zeeman coupling. Each layer of the unit cell contains two Cr atoms, $\tau\in\{\alpha,\beta\}$, related by the in-plane glide, and the bilayer unit cell hosts four magnetic sites labelled by the layer $\ell\in\{A,B\}$ and sublattice $\tau$. We collect the unit-cell position, layer, and sublattice into a single site index $i\equiv(\mathbf r_i,\ell_i,\tau_i)$, so that $\bS_i$ denotes the spin at site $i$. The model Hamiltonian is then
\begin{align}
\hat H
&= \sum_{\langle ij\rangle} J_{ij}\,\bS_{i}\!\cdot\!\bS_{j}
 - K\sum_{i}\big(S^{b}_{i}\big)^{2}
 - g_{e}\mu_{B}\,\mathbf B\!\cdot\!\sum_{i}\bS_{i},
\label{eq:Hbi}
\end{align}
with $S = 3/2$ per Cr$^{3+}$. The sum $\langle ij\rangle$ runs over all exchange bonds, and $J_{ij}$ takes the intralayer value $J^{\parallel}$ for bonds within a layer ($\ell_i=\ell_j$) and the interlayer value $J^{\perp}$ for bonds between layers ($\ell_i\neq\ell_j$). The glide symmetry enforces $J^{\parallel,\alpha\alpha}=J^{\parallel,\beta\beta}$ and $J^{\perp,\alpha\alpha}=J^{\perp,\beta\beta}$. Ferromagnetic intralayer exchanges ($J^{\parallel}<0$) align the moments along the easy axis selected by $K>0$, while the weaker antiferromagnetic interlayer exchanges ($J^{\perp}>0$) produce the A-type N\'{e}el stacking, with all values listed in Table~\ref{tab:Jvalues}. The anisotropy is written as $-K\sum_{i}(S^{b}_{i})^{2}$ to match the macroscopic easy axis $\hat b$ inferred from magnetization measurements and inelastic neutron scattering measurements.

\begin{table}[h!]
\centering
\caption{Heisenberg exchange constants. Intralayer values $J_{1\ldots 7}$
are from Inelastic Neutron Scattering data (Ref. \protect\citesi{https://doi.org/10.1002/advs.202202467-si})
classified following Ref. \protect\citesi{Bo_2023-si}; $J_6$ is the intralayer second-NN along $\pm 2\hat a$, not
out-of-plane. The interlayer $J_{z1},J_{z2}$ are from GGA$+U$ FPLR DFT.
The easy-axis spin-flip field $B_{\rm c}$ calibrates an
effective interlayer exchange $J_{\perp}^{\rm eff}$ via
Eq.~\eqref{eq:Bc-flip}. The saturation field $\Bsat$ then calibrates
an effective anisotropy $K$ via Eq.~\eqref{eq:canting}, in which
$\Bsat$ is set jointly by $K$ and by the interlayer molecular field
$2 J_{\perp}^{\rm eff} z_{\perp}$. Lattice constants
$a=3.50$\,\AA, $b=4.76$\,\AA, $c=7.96$\,\AA.}
\label{tab:Jvalues}
\renewcommand{\arraystretch}{1.25}
\setlength{\tabcolsep}{3pt}
\footnotesize
\begin{tabular}{c c c c l}
\hline\hline
Label & $J_{ij}$ (meV) & Bond vector $\hat d$ & coordination($z$) & Sublattice pair \\
\hline
\multicolumn{5}{c}{\textbf{Intralayer Heisenberg exchange}} \\
\hline
$J_1$ & $-1.90$  & $(\pm a,\,0)$              & 2 & $\alpha\!-\!\alpha$, $\beta\!-\!\beta$ \\
$J_2$ & $-3.38$  & $(\pm a/2,\,\pm b/2)$       & 4 & $\alpha\!-\!\beta$ \\
$J_3$ & $-1.67$  & $(0,\,\pm b)$              & 2 & $\alpha\!-\!\alpha$, $\beta\!-\!\beta$ \\
$J_4$ & $-0.09$  & $(\pm a,\,\pm b)$           & 4 & $\alpha\!-\!\alpha$, $\beta\!-\!\beta$ \\
$J_5$ & $-0.09$  & $(\pm 3a/2,\,\pm b/2)$      & 4 & $\alpha\!-\!\beta$ \\
$J_6$ & $+0.37$  & $(\pm 2a,\,0)$             & 2 & $\alpha\!-\!\alpha$, $\beta\!-\!\beta$ \\
$J_7$ & $-0.29$  & $(\pm a/2,\,\pm 3b/2)$      & 4 & $\alpha\!-\!\beta$ \\
\hline
\multicolumn{5}{c}{\textbf{Interlayer Heisenberg exchange}} \\
\hline
$J_{z1}$ & $+0.0008$ & diagonal, $6.62$\,\AA  & 4 & $\alpha\!-\!\beta$ (A/B) \\
$J_{z2}$ & $-0.0008$ & axial, $7.97$\,\AA    & 2 & $\alpha\!-\!\alpha$, $\beta\!-\!\beta$ (A/B) \\
\hline
\multicolumn{5}{c}{\textbf{Effective interlayer exchange from spin-flip calibration}} \\
\hline
$B_{\rm c}$ & $0.4$\,T & ($\mathbf B\!\parallel\!\hat b$) & 4 &
$J_{\perp}^{\rm eff} = g_{e}\mu_{B} B_{\rm c}/(2\,z_{\perp}S) \approx 3.9\;\mu$eV \\
\hline
\multicolumn{5}{c}{\textbf{Effective anisotropy from saturation-field calibration}} \\
\hline
$\Bsat$ & $2.1$\,T & ($\mathbf B\!\parallel\!\hat c$) & --- &
$K = g_{e}\mu_{B}\Bsat/(2S) - 2 J_{\perp}^{\rm eff} z_{\perp} \approx 50\;\mu$eV \\
\hline\hline
\end{tabular}
\end{table}

\subsection{Magnetic ordering temperatures}

Bulk CrSBr orders antiferromagnetically at $\TN\approx132$~K, yet strong intralayer ferromagnetic correlations set in only slightly higher, giving a Curie-like scale $\TC\approx150$~K. This near coincidence is striking because the interlayer exchange is smaller than the intralayer exchange by more than two orders of magnitude. To see why the two scales track each other, we reduce the bilayer model of Eq.~\eqref{eq:Hbi} to a minimal single-scale form on a square lattice, keeping one effective ferromagnetic intralayer exchange $|J^{\parallel}|$, one antiferromagnetic interlayer exchange $J^{\perp}$, and the easy-$\hat b$ anisotropy $K$. In the convention of Eq.~\eqref{eq:Hbi} the intralayer bonds are ferromagnetic ($J^{\parallel}<0$) and the interlayer bonds antiferromagnetic ($J^{\perp}>0$), so their magnitudes enter the estimates below. From Table~\ref{tab:Jvalues}, $|J^{\parallel}|\approx2.5$~meV, $K\approx50~\mu$eV, and $J^{\perp}\approx3.9~\mu$eV, with intralayer (interlayer) coordination $z_{\parallel}$ ($z_{\perp}$) and $S=3/2$.

At the mean-field level, let $m_l=\langle S^{b}_{i,l}\rangle$ be the easy-axis magnetization of layer $l$. Linearizing the mean-field equation gives
\begin{equation}\label{eq:MFlinear}
 m_l = \chi_{0,b}(T,K)\left[z_{\parallel}|J^{\parallel}|\,m_l - J^{\perp}\left(m_{l-1}+m_{l+1}\right)\right],
\end{equation}
where the bracket is the effective field in layer $l$ and $\chi_{0,b}$ is the longitudinal susceptibility of a single spin with the anisotropy $-K(S^{b})^{2}$,
\begin{equation}\label{eq:chi0}
 \chi_{0,b}(T,K) = \frac{1}{\kb T}\frac{\sum_{m=-S}^{S}m^{2}\exp[K m^{2}/(\kb T)]}{\sum_{m=-S}^{S}\exp[K m^{2}/(\kb T)]}.
\end{equation}
For ferromagnetic layers with antiferromagnetic stacking, $m_l=(-1)^l m$, and the antiferromagnetic and uniform ferromagnetic instabilities are set by
\begin{equation}\label{eq:MFtwo}
 1 = \left(z_{\parallel}|J^{\parallel}|\pm z_{\perp}J^{\perp}\right)\chi_{0,b}(T,K),
\end{equation}
with the $+$ ($-$) sign giving $\TN$ ($\TC$). For $K=0$, $\chi_{0,b}(T,0)=S(S+1)/3\kb T$, so
\begin{equation}\label{eq:TNMF}
 \kb\TN^{\rm MF} = \frac{S(S+1)}{3}\left(z_{\parallel}|J^{\parallel}|+z_{\perp}J^{\perp}\right).
\end{equation}
The two molecular fields in Eq.~\eqref{eq:MFtwo} differ only by the tiny term $2z_{\perp}J^{\perp}$, so $\TN$ and $\TC$ are nearly degenerate already here, both set by the intralayer exchange alone.

Mean-field theory misses the strong two-dimensional fluctuations that, by the Mermin-Wagner theorem, would forbid order in an isolated isotropic layer. A weak anisotropy restores it with only a logarithmic dependence on $K$. For small momenta the ferromagnetic magnon dispersion is $\epsilon_{\q}\approx\rho q^{2}+\Delta_{K}$, with spin-wave stiffness $\rho$ and anisotropy gap $\Delta_{K}$, and the thermal magnon population is
\begin{equation}\label{eq:magred}
 \delta M = \int\frac{d^{2}q}{(2\pi)^{2}}\frac{1}{\exp[(\rho q^{2}+\Delta_{K})/\kb T]-1}\approx\frac{\kb T}{4\pi\rho}\ln\!\left(\frac{\kb T}{\Delta_{K}}\right).
\end{equation}
Setting $\delta M\sim S$ and using $\kb\TC\sim|J^{\parallel}|$ with $\rho\sim|J^{\parallel}|S^{2}$ gives the leading-logarithmic form
\begin{equation}\label{eq:TC2D}
 \kb\TC \sim \frac{4\pi\rho}{\ln(|J^{\parallel}|/\Delta_{K})}\quad\Longrightarrow\quad \TC^{2D}\sim\frac{|J^{\parallel}|}{\ln(|J^{\parallel}|/K)},
\end{equation}
so $\TC$ vanishes only logarithmically as $K\to0$.

The interlayer coupling cuts off the two-dimensional fluctuations in a similar way. If layer $l+1$ develops a magnetization $m_{l+1}$, layer $l$ feels a field $h_l=-J^{\perp}m_{l+1}$ and responds with $m_l=\chi_{2D}h_l$, so with $m_l=(-1)^l m$ a finite moment requires
\begin{equation}\label{eq:RPAordering}
 z_{\perp}J^{\perp}\,\chi_{2D}(\TN)=1.
\end{equation}
For a 2D Heisenberg magnet the correlation length is exponential, $\xi_{2D}\sim\exp(2\pi\rho/\kb T)$, so that $\chi_{2D}\sim\xi_{2D}^{2}/\kb T$ and Eq.~\eqref{eq:RPAordering} becomes
\begin{equation}\label{eq:critJp}
 \frac{z_{\perp}J^{\perp}}{\kb T}\exp(2\pi\rho/\kb T)\sim1.
\end{equation}
Inverting, and using $\kb\TN\sim|J^{\parallel}|$ up to logarithmic corrections,
\begin{equation}\label{eq:TNJp}
 \kb\TN \sim \frac{4\pi\rho}{\ln(|J^{\parallel}|/J^{\perp})}.
\end{equation}
When both cutoffs are present they combine into an effective infrared scale, so that
\begin{equation}\label{eq:Tcombined}
 \kb\TN \sim \frac{4\pi\rho}{\ln(|J^{\parallel}|/\Delta_{\rm IR})},\qquad \Delta_{\rm IR}\approx c_{K}K+c_{\perp}J^{\perp},
\end{equation}
which follows the detailed quasi-2D Heisenberg analysis of Refs.~\citesi{IrkhinKatanin1997,IrkhinKatanin2000}.

The closeness of $\TC$ and $\TN$ now follows from Eqs.~\eqref{eq:TC2D}--\eqref{eq:Tcombined}. The intralayer correlations that produce the Curie-like feature and the interlayer locking that produces the N\'eel transition are both governed by the same large stiffness $\rho\sim|J^{\parallel}|S^{2}$, while the weak couplings appear only inside the logarithm. Moreover the cutoff $\Delta_{\rm IR}$ is dominated by the larger scale, the anisotropy $K$, since $J^{\perp}/K\approx0.08$, so adding the interlayer coupling shifts $\ln(|J^{\parallel}|/\Delta_{\rm IR})$ by only a few percent even though $J^{\perp}$ is what ultimately drives three-dimensional order. Both temperatures are therefore pinned near $4\pi\rho/\ln(|J^{\parallel}|/K)$, consistent with the observed $\TC\approx150$~K and $\TN\approx132$~K lying within a few tens of kelvin rather than the orders of magnitude one might naively expect from $J^{\perp}\ll|J^{\parallel}|$.

\section{Fleury--Loudon vertex}\label{SI_FL}

The Fleury--Loudon (FL) vertex describes how light couples to spins in a
magnetic insulator described by the following effective Hamiltonian,
\begin{align}
\hat H_{\rm FL} \;=\; \sum_{\langle ij\rangle}\,
\bigl(\hat{\boldsymbol\varepsilon}_{\rm in}\cdot\vec d_{ij}\bigr)\,
\bigl(\hat{\boldsymbol\varepsilon}_{\rm out}\cdot\vec d_{ij}\bigr)\,
J_{ij}\,\hat{\textbf {S}}_i \cdot \hat{\textbf {S}}_j,
\label{eq:HFL}
\end{align}
where $\hat{\boldsymbol\varepsilon}_{\rm in,out}$ are the incident and scattered
photon polarizations, $\vec d_{ij}$ is the bond vector between Cr sites $i$ and
$j$, and $J_{ij}$ is the Heisenberg exchange along that bond, with the sum over
all pairs. We evaluate this Hamiltonian using the Holstein--Primakoff (HP) transformation to leading order in $1/S$, using the linear spin wave theory (LSWT). Given that there are two layers in a unit cell of CrSBr in which the spins are anti-aligned along the quantization axis ($\pm\hat z$), we assign
independent bosons $\hat a^\dagger$ and $\hat b^\dagger$, to create magnons in layer A and in layer B, respectively.
\begin{align}
\text{layer A : } & \hat S^{z} \;=\; S - \hat a^{\dagger} \hat a, &
\hat S^{+} &\;\approx\; \sqrt{2S}\,\hat a, \\[4pt]
\text{layer B : } & \hat S^{z} \;=\; -\bigl(S - \hat b^{\dagger} \hat b\bigr), &
\hat S^{+} &\;\approx\; \sqrt{2S}\,\hat b^{\dagger} .
\end{align}
Decomposing $\hat{\textbf {S}}_i \cdot \hat{\textbf {S}}_j = \hat S^{z}_i\hat S^{z}_j + \tfrac12(\hat S^{+}_i\hat S^{-}_j + \hat S^{-}_i\hat S^{+}_j)$, followed by bosonization yields,
\begin{equation}
\begin{aligned}
\hat H_{\rm FL} \;=\;
&\phantom{+}\sum_{\langle ij\rangle\,\in\,{\rm intralayer}}\!\!\!\Lambda_{ij}\,J_{ij}
\Bigl[
\;-\;\underbrace{S\bigl(\hat a^{\dagger}_i\hat a_i + \hat a^{\dagger}_j\hat a_j\bigr)}_{\text{site-diagonal}}
\;+\;\underbrace{S\bigl(\hat a^{\dagger}_i\hat a_j + \hat a^{\dagger}_j\hat a_i\bigr)}_{\text{hopping (bilinear)}}
\Bigr] + (a \rightarrow b) \\[4pt]
&+\sum_{\langle ij\rangle\,\in\,{\rm interlayer}}\!\!\!\Lambda_{ij}\,J_{ij}
\Bigl[
\;+\;\underbrace{S\bigl(\hat a^{\dagger}_i\hat a_i + \hat b^{\dagger}_j\hat b_j\bigr)}_{\text{site-diagonal}}
\;+\;\underbrace{S\bigl(\hat a^{\dagger}_i\hat b^{\dagger}_j + \hat a_i\hat b_j\bigr)}_{\text{pair (bilinear)}}
\Bigr] \;+\;\mathcal O(1/S).
\end{aligned} \label{eq:bilinear}
\end{equation}
The constant term leaves the spins unchanged and the number conserving terms (site diagonal and intralayer hopping) leave the magnon count unchanged, so neither contributes to two magnon emission. Only the interlayer pair term does, taking the form $\hat a^{\dagger}_i\hat b^{\dagger}_j + \hat a_i\hat b_j$, which
changes the magnon number by two. Acting on the ground state,
$\hat a^{\dagger}_i\hat b^{\dagger}_j$ creates one magnon in layer A and one in
layer B, so the FL vertex emits a correlated magnon pair only across the interlayer bonds.

The operator responsible for two magnon scattering is therefore
\begin{align}
\hat R_{\rm FL} \;=\; S\!\!\sum_{\langle ij\rangle\,\in\,{\rm interlayer}}\!\!
\Lambda_{ij}\,J_{ij}\,\hat a^{\dagger}_i\hat b^{\dagger}_j ,
\label{eq:Rop}
\end{align}
built exclusively from the interlayer bonds and weighted by $J_{ij}$. We restrict
this sum to the $J_{z1}$ bonds (Fig.~\ref{fig:new_fig_1}(c)), which connect a Cr site in layer A to its neighbours in layer B at the in-plane displacements $\pm\hat{\mathbf d}_{\parallel}$. Fourier transforming the magnon operators, $\hat a^{\dagger}_i = N^{-1/2}\sum_{\bq} e^{-i\bq\cdot\mathbf r_i}\,\hat a^{\dagger}_{\bq}$ and likewise for $\hat b$, and noting that the photon transfers negligible momentum,
\begin{align}
\hat R_{\rm FL} \;=\; S\sum_{\bq} F(\bq)\,
\hat a^{\dagger}_{\bq}\hat b^{\dagger}_{-\bq},
\qquad
F(\bq)=\sum_{d\in J_{z1}}\Lambda_{d}\,J_{z1}\,e^{i\bq\cdot\hat{\mathbf d}},
\label{eq:Rq}
\end{align}
so the two emitted magnons carry opposite momenta $\bq$ and $-\bq$. The $J_{z1}$
bonds occur in centrosymmetric pairs $\pm\hat{\mathbf d}$, so the two phases in
each pair combine as $e^{i\bq\cdot\hat{\mathbf d}}+e^{-i\bq\cdot\hat{\mathbf d}}
=2\cos(\bq\cdot\hat{\mathbf d})$. The imaginary parts cancel, leaving a real
structure factor, and because $\bq$ lies in plane only the projection
$\hat{\mathbf d}_{\parallel}$ enters, so
$F(\bq)=2\sum_{\rm pairs}\Lambda_{d}\,J_{z1}\,\cos(\bq\cdot\hat{\mathbf d}_{\parallel})$.
A Bogoliubov transformation recasts $\hat a_{\bq}$ and $\hat b_{-\bq}$ into the
magnon eigenmodes $\nu_1,\nu_2$ with energies $\omega^{(\nu_1)}_{\bq}$ and
$\omega^{(\nu_2)}_{-\bq}$, so each emitted pair appears at the sum of its two
branch energies. Fermi's golden rule then yields the bimagnon lineshape
\begin{equation}
I_{\rm bm}(\omega) \;\propto\; \sum_{\bq}\,|F(\bq)|^{2}\,
\delta\!\big(\omega-\omega^{(\nu_{1})}_{\bq}-\omega^{(\nu_{2})}_{-\bq}\big),
\qquad
F(\bq) \;=\; \sum_{d\in J_{z1}}\Lambda_{d}\,J_{z1}\,
\cos(\bq\!\cdot\!\hat{\mathbf d}_{\parallel}).
\label{eq:Ibm}
\end{equation}
The spectral weight $|F(\bq)|^{2}$ is set entirely by the interlayer exchange
$J_{z1}$, so the integrated bimagnon intensity is a direct optical readout of the
interlayer spin correlation, tracking how strongly adjacent layers remain
anti-aligned and vanishing as the interlayer alignment is removed by field. We
stress that the quantity measured is the dynamical (four-spin) correlation
function of the interlayer bond operator, not the static two-spin correlator
$\langle\hat{\textbf S}_A\cdot\hat{\textbf S}_B\rangle$ nor the momentum-resolved
structure factor $S(\bq,\omega)$. Its weight and field dependence are
nonetheless set by, and therefore directly report on, the interlayer spin
correlation.

\section{Magnon density of states}\label{SI_twomagnon}

\begin{figure}[h!]
    \centering
    \includegraphics[width=\linewidth]{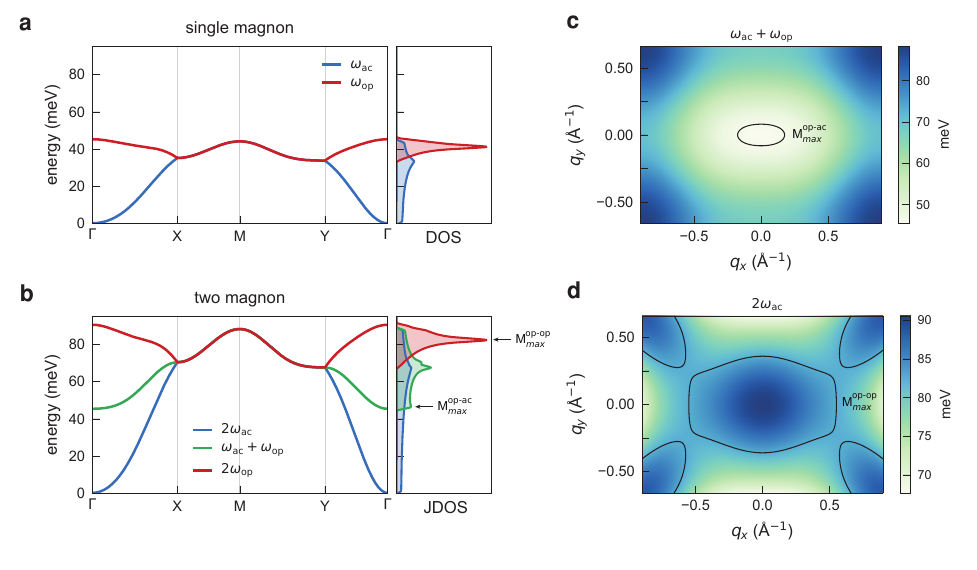}
    \caption{\textbf{Two-magnon joint density of states.}
    \textbf{a}, Single-magnon dispersion calculated from the Heisenberg model with the exchange-coupling parameters of Ref.~\citesi{https://doi.org/10.1002/advs.202202467-si}, comprising an acoustic and an optical branch.
    \textbf{b}, Two-magnon dispersion obtained by pairing magnons of equal and opposite momentum, yielding three branches from the combinations of the two single-magnon branches.
    \textbf{c}, Weight of the $\rm ac+op$ channel at the band-edge feature $M^{\rm op-ac}_{\rm max}\!\approx\!46.2$~meV, showing a closed loop around $\Gamma$ where the optical-magnon gap sets the two-magnon threshold.
    \textbf{d}, Weight of the $\rm op+op$ channel at the dominant peak $M^{\rm op-op}_{\rm max}\!\approx\!82.5$~meV, showing a saddle-point contour threading the $M$ corners of the Brillouin zone that produces the logarithmic divergence of the two-magnon density of states $\rho^{(2)}(\omega)$ at $2\omega^{\rm op}_{\rm max}$.}
    \label{fig:SI_twomagnon}
\end{figure}

The single-magnon dispersion of Fig.~\ref{fig:SI_twomagnon}(a) comprises an acoustic and an optical branch. Pairing magnons of equal and opposite momentum builds the non-interacting two-magnon dispersion, whose three branches arise from the combinations of the two single-magnon branches (Fig.~\ref{fig:SI_twomagnon}(b)). The resulting two-magnon density of states exhibits Van Hove singularities that coincide with the spectral features of the Raman spectra, and the Brillouin-zone regions from which they originate are mapped in Fig.~\ref{fig:SI_twomagnon}(c,d).

\section{Magnon-magnon interactions and the $M^{\rm op-op}_{\rm max}$ shift}\label{SI_twomagnon_interaction}

The bare Fleury-Loudon lineshape of Eq.~\eqref{eq:Ibm} reproduces both measured two-magnon peaks, except that the optical-optical feature $M^{\rm op-op}_{\rm max}$ lies about $10~\mathrm{meV}$ above the calculated optical-optical continuum. We now show that this residual shift is naturally accounted for by magnon-magnon interactions, treated as a ladder (RPA) resummation of the two-magnon susceptibility, and that it is channel-selective, appearing in the nearly flat optical-optical sector while leaving the dispersive acoustic-optical features unshifted. Because the magnon bands are set by the intralayer ferromagnetic exchange, for which the fully polarized state and its single-magnon excitations are exact eigenstates, the linear spin-wave dispersion carries no $1/S$ correction, so the residual shift necessarily originates in the two-magnon sector as a magnon-magnon interaction rather than a single-magnon renormalization.

The starting point is the Fleury-Loudon operator $\hat R_{\rm FL}$ introduced above. After the spin-wave expansion, the Bogoliubov transformation recasts the layer bosons $\hat a_{\bq},\hat b_{-\bq}$ into the magnon eigenmodes $\hat\beta_{\nu,\bq}$ with $\nu\in\{a,o\}$ labelling the acoustic and optical branches. Restricted to the optical branch, the pair-creation part of $\hat R_{\rm FL}$ is
\begin{equation}\label{eq:RamanPair}
 \hat R_{oo} = \sum_{\bq}F_{oo}(\bq)\left(\hat\beta^{\dagger}_{o,\bq}\hat\beta^{\dagger}_{o,-\bq} + \hat\beta_{o,-\bq}\hat\beta_{o,\bq}\right),
\end{equation}
where $F_{oo}(\bq)$ is the optical-optical projection of the interlayer structure factor $F(\bq)$ of Eq.~\eqref{eq:Ibm}, dressed by the Bogoliubov coefficients that define the eigenmodes. The retarded Raman susceptibility $\chi^{R}(t)=-i\theta(t)\langle[\hat R_{oo}(t),\hat R_{oo}(0)]\rangle$ evaluates, at the linear spin-wave level, to the bare two-magnon propagator
\begin{equation}\label{eq:Pi0raw}
 \chi^{R}(\Omega) = \sum_{\bq}\frac{|F_{oo}(\bq)|^{2}}{\Omega - 2\omega^{(o)}_{\bq} + i\eta},
\end{equation}
where $\Omega$ is the Raman shift and $\omega^{(o)}_{\bq}$ the optical-branch magnon energy. Its imaginary part $I_0(\Omega)\propto-\im\chi^{R}(\Omega)=\sum_{\bq}|F_{oo}(\bq)|^{2}\delta(\Omega-2\omega^{(o)}_{\bq})$ is precisely the vertex-weighted two-magnon density of states of Eq.~\eqref{eq:Ibm}. Defining a normalized pair wavefunction $\phi_{\bq}=F_{oo}(\bq)/\sqrt{\sum_{\mathbf k}|F_{oo}(\mathbf k)|^{2}}$ with $\sum_{\bq}|\phi_{\bq}|^{2}=1$, we write the normalized bare propagator
\begin{equation}\label{eq:Pinorm}
 \Pi_0(\Omega) = \sum_{\bq}\frac{|\phi_{\bq}|^{2}}{\Omega - E_{\bq} + i\eta},\qquad E_{\bq}=2\omega^{(o)}_{\bq},
\end{equation}
so that $\Pi_0$ has units of inverse energy and the projected interaction defined below has units of energy.

Magnon-magnon interactions arise from the quartic terms of the spin-wave Hamiltonian. Continuing the expansion to that order defines an effective interaction in the optical-optical channel, $H_4^{oo}=\sum_{\bq,\bq'}V_{oooo}(\bq,\bq')|\bq\rangle\langle\bq'|$, where $|\bq\rangle=\hat\beta^{\dagger}_{o,\bq}\hat\beta^{\dagger}_{o,-\bq}|0\rangle$ and $V_{oooo}$ is the four-magnon vertex projected onto the optical branch. Since the Raman operator creates the particular normalized superposition $|R\rangle=\hat R_{oo}|0\rangle/\sqrt{\langle0|\hat R_{oo}^{\dagger}\hat R_{oo}|0\rangle}=\sum_{\bq}\phi_{\bq}|\bq\rangle$, the physically relevant coupling is its projection onto the Raman-active state,
\begin{equation}\label{eq:URmicro}
 U_R = \langle R|H_4^{oo}|R\rangle = \sum_{\bq,\bq'}\phi_{\bq}^{*}V_{oooo}(\bq,\bq')\phi_{\bq'}.
\end{equation}
Approximating the interaction as separable within this subspace, $V(\bq,\bq')=U_R\,\phi_{\bq}\phi_{\bq'}^{*}$, or equivalently $H_4^{oo}=U_R|R\rangle\langle R|$, is well justified for a nearly flat band, since when $E_{\bq}\approx E_0$ the bare propagator $(\Omega-E_{\bq}+i\eta)^{-1}$ is nearly momentum independent and free propagation does not reshape the Raman-created pair. The ladder series $\Pi_0+\Pi_0 U_R\Pi_0+\Pi_0 U_R\Pi_0 U_R\Pi_0+\cdots$ then resums to
\begin{equation}\label{eq:RPA}
 \chi_R(\Omega) = \frac{\Pi_0(\Omega)}{1 - U_R\,\Pi_0(\Omega)},
\end{equation}
which is the standard ladder approximation to two-magnon Raman scattering~\citesi{CanaliGirvin1992,Liu2017}. With the convention that $U_R>0$ is repulsive, an antibound state appears above the bare continuum when
\begin{equation}\label{eq:pole}
 1 - U_R\,\re\Pi_0(\Omega_B) = 0,\qquad \Omega_B > E_{\max}\equiv\max_{\bq}E_{\bq}.
\end{equation}
Because $\im\Pi_0(\Omega)=0$ for $\Omega>E_{\max}$ while $\re\Pi_0$ stays finite, the denominator of Eq.~\eqref{eq:RPA} can vanish outside the bare continuum, producing a genuine two-particle pole at an energy where the bare two-magnon density of states is zero. This antibound magnon pair, pushed above the continuum by the repulsive interaction, is the origin of the upward shift.

The size of the shift follows from $\Pi_0$. In the perfectly flat limit $\omega^{(o)}_{\bq}=\omega_o$, every pair has energy $E_0=2\omega_o$, so $\Pi_0(\Omega)=(\Omega-E_0+i\eta)^{-1}$ and Eq.~\eqref{eq:RPA} gives
\begin{equation}\label{eq:flatRPA}
 \chi_R(\Omega) = \frac{1}{\Omega - E_0 - U_R + i\eta},
\end{equation}
a single pole at $\Omega_B=2\omega_o+U_R$, so the interaction-induced shift is simply $\Delta=U_R$. A measured optical-optical peak lying $\approx10~\mathrm{meV}$ above the bare optical-optical energy therefore corresponds, at crudest level, to $U_R\approx10~\mathrm{meV}$. A small but finite bandwidth $W$ refines this. Approximating the two-magnon density of states by a box $\rho_2(E)=1/W$ on $E_0<E<E_0+W$, the real part of the propagator at $\Omega_B=E_0+W+\delta$ with $\delta>0$ is
\begin{equation}\label{eq:boxPi}
 \re\Pi_0(\Omega_B) = \frac{1}{W}\int_{E_0}^{E_0+W}\frac{dE}{\Omega_B-E} = \frac{1}{W}\ln\!\left(1+\frac{W}{\delta}\right),
\end{equation}
so the pole condition gives $U_R=W/\ln(1+W/\delta)$, which expands to $U_R\approx\delta+W/2$ and recovers the flat-band result as $W\to0$. For a peak $\delta\approx10~\mathrm{meV}$ above the bare edge, $W=5~\mathrm{meV}$ gives $U_R=12.3~\mathrm{meV}$ and $W=10~\mathrm{meV}$ gives $U_R=14.4~\mathrm{meV}$, so $U_R\approx10\!-\!15~\mathrm{meV}$ for a moderately narrow optical band.

A sharper estimate avoids the box model entirely and uses the bare Fleury-Loudon dispersion directly. Normalizing $\phi_{\bq}=F_{oo}(\bq)/\sqrt{\sum_{\mathbf k}|F_{oo}(\mathbf k)|^{2}}$ and evaluating $\Pi_0(\Omega)=\sum_{\bq}|\phi_{\bq}|^{2}/(\Omega-2\omega^{(o)}_{\bq}+i\eta)$, the interaction that shifts the calculated peak onto the observed value $\Omega_{\rm exp}$ is
\begin{equation}\label{eq:URexact}
 U_R = \frac{1}{\re\Pi_0(\Omega_{\rm exp})},
\end{equation}
and the corresponding RPA spectrum is
\begin{equation}\label{eq:IRRPA}
 I_R(\Omega)\propto-\im\frac{\Pi_0(\Omega)}{1-U_R\Pi_0(\Omega)}.
\end{equation}
This automatically incorporates the full magnon dispersion and the momentum dependence of the Fleury-Loudon vertex, and is therefore more accurate than referencing the shift to twice the zone-center optical energy.

Finally, the same construction explains why only the optical-optical peak shifts. For an acoustic-optical pair the bare propagator is $\Pi_{ao}^{(0)}(\Omega)=\sum_{\bq}|F_{ao}(\bq)|^{2}/(\Omega-\omega^{(a)}_{\bq}-\omega^{(o)}_{-\bq}+i\eta)$, and the projected interaction $U_{\mu\nu}=\sum_{\bq,\bq'}\phi_{\mu\nu,\bq}^{*}V_{\mu\nu\mu\nu}(\bq,\bq')\phi_{\mu\nu,\bq'}$ has no reason to equal $U_{oo}$, since the eigenvectors, bare vertex, and two-magnon density of states all differ between channels. The nearly flat optical branch is especially favorable for interaction effects, because its large two-particle density of states lets repeated scattering build a well-defined antibound state above the continuum, whereas the broad acoustic-optical continuum yields a much smaller shift even though both quartic interactions originate from the same spin Hamiltonian. Consequently, the fact that linear spin-wave theory reproduces the acoustic-optical features while its optical-optical continuum terminates $\approx10~\mathrm{meV}$ below the measured $M^{\rm op-op}_{\rm max}$ peak is strong evidence for a channel-selective magnon-magnon interaction concentrated in the optical-optical sector.

\section{Polarization angle dependence of Raman emission}\label{SI_pol}
\begin{figure}[h!]
    \centering
    \includegraphics[width=\linewidth]{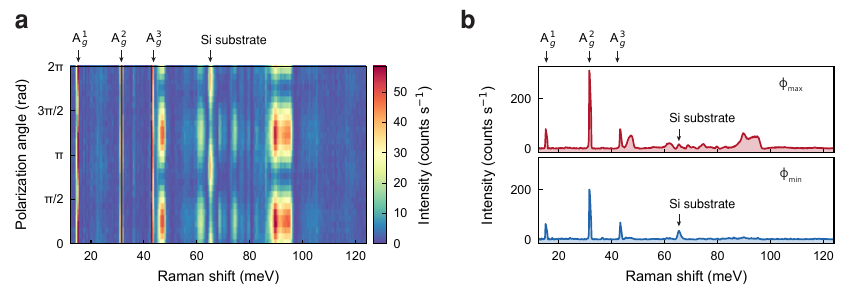}
    \caption{\textbf{Polarization dependence of Raman Spectra (electronic-band mediated).}
    \textbf{a}, A 2D map representing the polarization dependence of the Raman spectra. \textbf{b}, Raman spectra plotted at $\phi_\text{max}$ and at $\phi_\text{min}$. Measurements were made with $P = 100$ $\mu$W and at $T=4$ K.}
    \label{fig:SI_pol}
\end{figure}

The sample was excited with linearly polarized light, and the in-plane
polarization was rotated. Because the Raman response is $\pi$-periodic in the
physical polarization angle, we plot it against $\phi$, defined as twice the
physical polarization angle, so that the response spans a full $0$ to $2\pi$;
thus $\phi=0$ corresponds to the crystalline $a$ axis and $\phi=\pi$ to the
crystalline $b$ axis. $\phi_{\rm max}$
($\phi_{\rm min}$) denotes the angle at which the two-magnon Raman intensity is
highest (lowest). The well-known Si substrate Raman peak at 64~meV
(521~cm$^{-1}$) is labeled. The Si peak is strongest when the polarization is
out of phase with $\phi_{\rm max}$, as there, less light is absorbed by CrSBr, so
more is transmitted to the substrate.

\section{Magnetic ground state and excitonic response}\label{SI_PL}

\begin{figure}[h!]
    \centering
    \includegraphics[width=\linewidth]{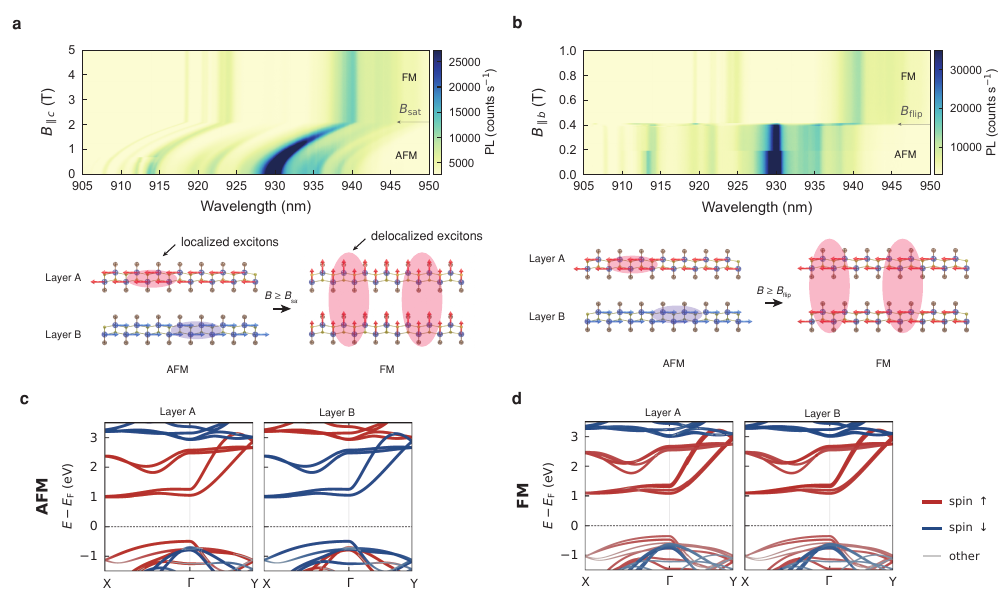}
    \caption{\textbf{Spin-flip behavior of CrSBr.}
    \textbf{a}, Top: photoluminescence spectra as a function of hard-axis magnetic field $\mathbf{B} \parallel \hat c$, taken at $T = 3.5$~K and $P = 100\,\mu$W. Bottom: schematic of the two adjacent CrSBr layers in the A-type AFM ground state at $B < B_{\rm sat}$ (canted precession) and in the field-polarized FM state at $B > B_{\rm sat}$.
    \textbf{b}, Top: photoluminescence spectra as a function of easy-axis magnetic field $\mathbf{B} \parallel \hat b$, taken at $T = 3.5$~K and $P = 100\,\mu$W. Bottom: schematic of the AFM ground state at $B < B_{\rm flip}$ and the field-polarized FM state at $B > B_{\rm flip}$ reached through the first-order spin-flip transition.
    \textbf{c}, Electronic band structure of layers $A$ and $B$ in the AFM ground state and in the field-polarized FM state, computed within DFT+$U$ and post-processed to a Wannier basis.}
    \label{fig:SI_PL}
\end{figure}

Fig.~\ref{fig:SI_PL}(a,b) shows the photoluminescence (PL) spectrum of bulk CrSBr as a function of magnetic field along the hard ($\mathbf{B}\!\parallel\!\hat c$) and easy ($\mathbf{B}\!\parallel\!\hat b$) axes, excited at $\lambda\!=\!633$~nm with $\hat{\mathbf{e}}\!\parallel\!\hat b$. The zero-field spectrum contains free excitons, exciton--polaritons~\citesi{Dirnberger2023-si,Wang2023-si}, and phonon side-bands~\citesi{Lin2024-si}, all of which red-shift monotonically under $\mathbf{B}\!\parallel\!\hat c$~\citesi{Wilson2021-si,Bae2022-si} and jump discontinuously at the metamagnetic threshold $B_{\rm flip}\!\approx\!0.4$~T under $\mathbf{B}\!\parallel\!\hat b$, tracking the smooth-canting and spin-flip trajectories of $\langle\hat{\mathbf S}_{A}\!\cdot\!\hat{\mathbf S}_{B}\rangle$ respectively.

The red-shift originates in the magnetic-state dependence of interlayer tunnelling. In the AFM ground state, the antiparallel spin alignment of adjacent layers blocks single-particle hopping at first order and confines the exciton to one layer; the spin-resolved fat-band calculation of Fig.~\ref{fig:SI_PL}(c) shows the band-edge states of layers $A$ and $B$ carrying opposite spin, with the nearest same-spin state several eV away. Continuous canting under $\mathbf{B}\!\parallel\!\hat c$ or the discontinuous spin flip under $\mathbf{B}\!\parallel\!\hat b$ progressively lifts this block; above $\Bsat$ or $B_{\rm flip}$ both layers acquire the same band-edge spin character, and the exciton delocalizes across the bilayer with reduced binding energy and lower emission energy.

\section{Magnetic field dependence of the magnon dispersion.}\label{SI_magnonBfield}

In this section, we quantify the magnon anisotropy ($K$) and the effective interlayer exchange ($J_\text{eff}$) from the $B_{\rm flip}$ and $B_{\rm sat}$ values experimentally obtained by PL measurements (Fig.~\ref{fig:SI_PL}).

\subsection{The effective interlayer coupling ($J_\perp$)}
An easy-axis field $\mathbf B\!\parallel\!\hat b$ exerts no Zeeman torque on the collinear AFM ground state, so no continuous canting of the moments develops. The moments stay collinear with $\hat b$ up to a threshold field, at which the fully polarized FM configuration becomes lower in energy than the AFM state ($E_{\rm FM}(B) \leq E_{\rm AFM}$). With the moments held along $\hat b$, the classical energy per formula unit of each state follows from Eq.~\eqref{eq:Hbi}, such that
\begin{align}
E_{\rm FM}(\Bc) &= E_{\rm AFM}, \nonumber\\
\underbrace{E_{\parallel}-KS^{2}}_{\text{common}}
 + J_{\perp}^{\rm eff} z_{\perp} S^{2}
 - g_{e}\mu_{B}\,\Bc S
&=
\underbrace{E_{\parallel}-KS^{2}}_{\text{common}}
 - J_{\perp}^{\rm eff} z_{\perp} S^{2} ,
\label{eq:Eflip}
\end{align}
with the intralayer exchange energy
$E_{\parallel}=S^{2}\sum_{\langle ij\rangle^{\parallel}} J^{\parallel}_{ij}$ and $z_{\perp}$ the interlayer coordination number.
The common intralayer and single-ion terms cancel from both sides, and the
interlayer exchange changes sign between the states while only the FM state
carries the Zeeman term, leaving
\begin{equation}
2 J_{\perp}^{\rm eff} z_{\perp} S^{2}
=  g_{e}\mu_{B}\,\Bc S
\quad\Longrightarrow\quad
\Bc = \frac{2J_{\perp}^{\rm eff}\,z_{\perp}\,S}{g_{e}\mu_{B}} .
\label{eq:Bc-flip}
\end{equation}
Inverting Eq.~\eqref{eq:Bc-flip} gives the effective interlayer exchange
from the measured $\Bc\!\approx\!0.4$~T:
\begin{equation}
J_{\perp}^{\rm eff} \;=\; \frac{g_{e}\mu_{B}\,\Bc}{2\,z_{\perp}\,S}
\;\approx\; \frac{(1.158\times 10^{-4}~\text{eV\,T}^{-1})(0.4~\text{T})}{4\times 3}
\;\approx\; 3.9\;\mu\text{eV},
\label{eq:Jperp-value}
\end{equation}
i.e.\ $J_{\perp}^{\rm eff}\,z_{\perp}\!\approx\!15.4~\mu$eV per Cr, roughly an order
of magnitude larger than the bare $J_{z1}$ obtained from DFT calculations
(Table~\ref{tab:Jvalues}). We therefore treat $J_{\perp}^{\rm eff}$ as a
phenomenological interlayer exchange that folds in interlayer superexchange paths and
correlation effects not captured by the bare GGA$+U$ diagonal. Only the bond geometry that
fixes the two-magnon selection rule, not the absolute magnitude of the interlayer exchange,
enters the Raman interpretation.

\subsection{The magnon anisotropy ($K$)}
A field $\mathbf B\!\parallel\!\hat c$ cants the sublattice moments from the
easy axis $\hat b$ toward $\hat c$ through an angle $\theta(B)$, so that the
local quantization axes become
\begin{equation}
\hat n_{A,B}(\theta)=(0,\pm\cos\theta,\sin\theta).
\label{eq:nhat}
\end{equation}

Introducing Holstein--Primakoff bosons about the local axes
Eq.~\eqref{eq:nhat}, the terms describing a uniform rigid tilt of the two
sublattices organize the Hamiltonian in Eq.~\eqref{eq:Hbi} as
$\epsilon(\theta)+\epsilon'(\theta)\,\delta\hat\theta
+\tfrac{1}{2}\epsilon''(\theta)\,\delta\hat\theta^{\,2}$ per site, where
$\delta\hat\theta\sim i\big[(a-a^{\dagger})-(b-b^{\dagger})\big]$ is the
magnon quadrature that rotates both sublattice axes, and $\epsilon(\theta)$ is the classical energy per site given explicitly below. Completing the square
with $\widetilde{\delta\theta}=\delta\hat\theta+\epsilon'/\epsilon''$ gives
$\tfrac{1}{2}\epsilon''\,\widetilde{\delta\theta}^{\,2}-\epsilon'^{2}/2\epsilon''$,
so a nonvanishing $\epsilon'$ merely displaces the quadrature, i.e.\
redefines the quantization axis, revealing that the expansion point was not
the energy minimum. Setting $\epsilon'(\theta)=0$ removes this shift and
leaves a positive semidefinite quadratic form, identifying the canting angle
as the configuration about which the magnon expansion should be performed. Substituting these axes into Eq.~\eqref{eq:Hbi}, the $\theta$-dependent classical
energy per site is
\begin{equation}
\epsilon (\theta) = \frac{E(\theta)}{N}
= -J_{\perp}z_{\perp}S^{2}\cos 2\theta
  \;-\; KS^{2}\cos^{2}\theta
  \;-\; g_{e}\mu_{B}BS\sin\theta ,
\label{eq:Eclass}
\end{equation}
where $N$ is the number of sites, the first term is the interlayer exchange
($\hat n_{A}\!\cdot\!\hat n_{B}=-\cos 2\theta$), the second the single-ion
easy-axis anisotropy $K$ along $\hat b$, and the third the Zeeman coupling to
the field along $\hat c$; the $\theta$-independent intralayer ferromagnetic
energy is constant and has been dropped. Minimizing with respect to $\theta$
($d\epsilon/d\theta=0$) yields
\begin{equation}
\sin\theta(B) = \frac{g_{e}\mu_{B}B}{2\,S\,(2J_{\perp}z_{\perp}+K)}
\equiv \frac{B}{\Bsat},\qquad
\Bsat=\frac{2\,S\,(2J_{\perp}z_{\perp}+K)}{g_{e}\mu_{B}},
\label{eq:canting}
\end{equation}
where the effective product
$J_{\perp}z_{\perp}\equiv\sum_{n}z_{\perp}^{(n)}\,J_{\perp}^{(n)}$
collapses the interlayer block of Table~\ref{tab:Jvalues} into a single
coupling. Inverting Eq.~\eqref{eq:canting} for $K$, with $J_{\perp}^{\rm eff}$ fixed in the
previous subsection and the measured $\Bsat\!\approx\!2.1$~T, gives
\begin{equation}
K = \frac{g_{e}\mu_{B}\,\Bsat}{2S} - 2 J_{\perp}^{\rm eff} z_{\perp}
\;\approx\; 81~\mu\text{eV} - 31~\mu\text{eV}
\;\approx\; 50~\mu\text{eV}.
\label{eq:K-value}
\end{equation}

Hence, across the entire measured range the magnon spectrum is essentially rigid,
because the field couples to the spins only through the Zeeman energy
$g_{e}\mu_{B}B$, which reaches at most $\sim\!0.1$--$0.2$~meV up to the
saturation field. This is two to three orders of magnitude smaller than the
magnon bandwidth and the optical gap ($\sim\!45$~meV), both set by the strong
intralayer exchange, which the field does not touch. The field therefore only
rescales the sub-meV anisotropy gap of the acoustic branch and leaves the
dispersion that governs the two-magnon energies unchanged, so the two-magnon peak
position is fixed and the observed field dependence is carried entirely by the
Loudon--Fleury vertex rather than by any shift of the magnon bands.

\section{Canted magnetic field geometries}\label{SI_Bfield}

\subsection{Spin-canting effect on Raman intensity}
Sec.~\ref{SI_magnon} treats the collinear case in which the moments lie along
$\pm\hat z$. A magnetic field tilts the two sublattice moments away from the
easy axis $\hat b$ by an angle $\theta(B)$ set by the field direction. Each
sublattice is again described in its own frame, with the local $z$ axis along
its moment $\hat n_{\ell}(\theta)$ and the local $x$ axis along $\hat a$, so
that both sublattices remain collinear in their own frame and the HP assignment
of Sec.~\ref{SI_FL} carries over unchanged. For $\mathbf B\!\parallel\!\hat c$
the moments tilt within the $bc$ plane,
$\hat n_{A,B}(\theta)=(0,\pm\cos\theta,\sin\theta)$, and the spins expressed in
the crystal frame $(\hat a,\hat b,\hat c)$ read
\begin{align}
\hat{\textbf S}_{A} &=
\begin{pmatrix}
\hat S^{x}_{A} \\[2pt]
\sin\theta\,\hat S^{y}_{A}+\cos\theta\,\hat S^{z}_{A} \\[2pt]
-\cos\theta\,\hat S^{y}_{A}+\sin\theta\,\hat S^{z}_{A}
\end{pmatrix},
&
\hat{\textbf S}_{B} &=
\begin{pmatrix}
\hat S^{x}_{B} \\[2pt]
\sin\theta\,\hat S^{y}_{B}-\cos\theta\,\hat S^{z}_{B} \\[2pt]
\cos\theta\,\hat S^{y}_{B}+\sin\theta\,\hat S^{z}_{B}.
\end{pmatrix},
\label{eq:rot}
\end{align}

Substituting Eq.~\eqref{eq:rot} into Eq.~\eqref{eq:HFL} and repeating the
bosonization, the interlayer bond is expressed as
\begin{align}
\hat{\textbf S}_{A}\!\cdot\!\hat{\textbf S}_{B}
={}& -\cos 2\theta\,S^{2} \nonumber\\[2pt]
&+S\cos^{2}\!\theta\,\bigl(\hat a^{\dagger}\hat b^{\dagger}+\hat a\hat b\bigr) \nonumber\\[2pt]
&+S\sin^{2}\!\theta\,\bigl(\hat a^{\dagger}\hat b+\hat a\hat b^{\dagger}\bigr) \nonumber\\[2pt]
&+\sqrt{\tfrac{S^{3}}{2}}\,\sin 2\theta\,
\bigl(\hat a+\hat a^{\dagger}-\hat b-\hat b^{\dagger}\bigr)
\;+\;\mathcal{O}(1/S).
\label{eq:dotcant}
\end{align}
Evidently, only the second line emits a pair. The $\hat a^{\dagger}\hat b^{\dagger}$
operator raises the magnon number by two, creating one magnon in each layer,
and is the sole two magnon channel. The remaining terms are inactive for this
process. The constant scatters elastically, the $\sin^{2}\!\theta$ term is
number conserving and only transfers a magnon between layers, and the
$\sin 2\theta$ term is linear in the boson operators and switches on the one
magnon scattering that is forbidden at $\theta=0$. Canting therefore leaves the
structure of the vertex intact and changes only the weight carried by each
term. Two magnon emission is still governed by the pair term of
Eq.~\eqref{eq:Rop}, now carrying an extra factor $\cos^{2}\!\theta$, so
Eqs.~\eqref{eq:Rq} and \eqref{eq:Ibm} hold under
$F(\bq)\to\cos^{2}\!\theta\,F(\bq)$ and
\begin{align}
I_{\rm bm}(\omega,B) \;\propto\; \cos^{4}\!\theta(B)\sum_{\bq}|F(\bq)|^{2}\,
\delta\!\big(\omega-\omega^{(\nu_{1})}_{\bq}-\omega^{(\nu_{2})}_{-\bq}\big).
\label{eq:IbmB}
\end{align}

The field enters only through $\theta(B)$. For a canted antiferromagnet the
equilibrium angle follows $\sin\theta = B/\Bsat$ up to the saturation field
$\Bsat$, at which the moments are fully polarized along the field. The
two-magnon amplitude is set by the pair-creation part of the interlayer
coupling, which measures how anti-aligned the two layers remain. For moments at
a relative angle $\varphi$ this pair-creation weight is $\sin^{2}(\varphi/2)$,
maximal when the layers are antiparallel and vanishing once they are parallel.
Because the spins in adjacent layers cant symmetrically from opposite in-plane
directions, the angle between them is $\varphi = \pi - 2\theta$, so the weight
becomes $\sin^{2}\!\bigl(\tfrac{\pi}{2}-\theta\bigr) = \cos^{2}\theta$. With
$\sin\theta = B/\Bsat$ this gives $\cos^{2}\theta = 1-(B/\Bsat)^{2}$, and the
integrated bimagnon intensity, set by the squared amplitude, falls off as
\begin{align}
I_{\rm bm}(B) \;\propto\; \cos^{4}\theta \;=\; \Bigl[1-(B/\Bsat)^{2}\Bigr]^{2},
\label{eq:IbmBsat}
\end{align}
vanishing at $\Bsat$, where the moments become parallel and the pair-creation
channel closes.

\subsection{Spin-flipping effect on Raman intensity}
For $B<\Bc$ the N\'{e}el state is unperturbed ($\cos 2\theta=1$) and the bimagnon matrix element is field-independent; for $B>\Bc$ the ground state is the saturated FM and the Heisenberg interlayer pair-creation operator in Eq.~\eqref{eq:bilinear} reduces to hopping, annihilating the magnon vacuum. The integrated bimagnon intensity is therefore a step function,
\begin{equation}
I_{\rm bm}(B) \;=\; I_{0}\,\Theta\!\big(\Bc-B\big) \;+\; I_{\rm SOC}\,\Theta\!\big(B-\Bc\big),
\label{eq:step-Bparb}
\end{equation}
with $I_{0}$ the zero-field plateau and $I_{\rm SOC}\!\ll\!I_{0}$ the residual SOC-mediated channel surviving in the polarized phase.

\section{Power dependence of the Raman spectrum}\label{SI_power}

\begin{figure}[h!]
    \centering
    \includegraphics[width=\linewidth]{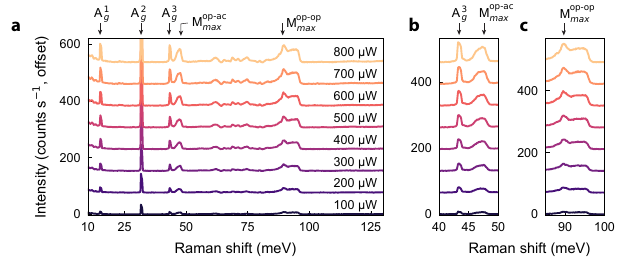}
    \caption{\textbf{Power dependence of the Raman spectrum (electronic-band mediated).}
    \textbf{a}, Waterfall of Raman spectra at $T=4$~K for laser powers from
    $100$ to $800~\mu$W in $100~\mu$W steps, excited with the
    $\lambda_0 = 633$~nm laser. \textbf{b,c}, Zoomed-in panels of the
    magnon-related peaks.}
    \label{fig:SI_power_1}
\end{figure}

\begin{figure}[h!]
    \centering
    \includegraphics[width=\linewidth]{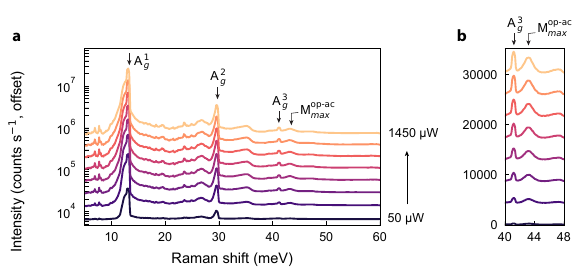}
    \caption{\textbf{Power dependence of the Raman spectrum (exciton mediated).}
    \textbf{a}, Waterfall of Raman spectra at $T=4$~K for laser powers from
    $50$ to $1450~\mu$W in $200~\mu$W steps, excited with the
    $\lambda_0 = 908$~nm laser. \textbf{b}, Zoomed-in panel of the
    magnon-related peaks.}
    \label{fig:SI_power_2}
\end{figure}

The power dependence of the Raman spectrum with different excitation regimes are shown in Figs.~\ref{fig:SI_power_1} and \ref{fig:SI_power_2}. There are no observable shifts in the Raman peaks with increasing power within the studied power regime.

\pagebreak

\makeatletter
\setcounter{NAT@ctr}{0}
\makeatother
\bibliographystylesi{sn-mathphys-num}
\bibliographysi{sn-bibliography-si}

\end{document}